\documentclass[pra,superscriptaddress,groupedaddress,twocolumn]{revtex4-1}
\usepackage{amssymb}
\usepackage{amsmath}
\usepackage{graphicx}
\usepackage{subfigure}	 	
\usepackage[english]{babel}
\usepackage{float}
\usepackage{color}
\usepackage{bm}
\usepackage{epsfig}

\begin{document}

\title{Multiplicity of vortex soliton families in the discrete
Ginzburg-Landau equation,\\their interactions and the formation of bound states}
\author{C. Mej\'ia-Cort\'es}
\author{J.M. Soto-Crespo}
\affiliation{Instituto de \'Optica, C.S.I.C., Serrano 121, 28006 Madrid, Spain}
\author{Rodrigo A. Vicencio}
\author{Mario I. Molina}
\affiliation{Departamento de F\'isica, Facultad de Ciencias, Universidad de
Chile, Casilla 653, Santiago, Chile}
\affiliation{Center for Optics and Photonics (CEFOP), Casilla 4016, Concepci\'on, Chile}
\date{\today}

\begin{abstract}
By using different continuation methods, we unveil a wide region in
the parameter space of the discrete cubic-quintic complex
Ginzburg-Landau equation, where several families of stable vortex
solitons coexist. All these stationary solutions have a symmetric amplitude
profile and two different topological charges. We also discover the dynamical 
formation of a variety of `bound-state' solutions, composed of two or more of
these vortex solitons. All of these stable composite structures persist in the
conservative cubic limit, 
for high values of their power content.

\end{abstract}


\maketitle
\section{Introduction}

Optical beams whose phase circulates around a singular point -or
central core-, changing by $2\pi S$ times in each closed loop around
it (with $S$ being an integer number), are called optical vortices.
The integer number $S$ is known as the topological charge of the
vortex, and its sign defines the direction of the phase circulation.
Usually an optical vortex has a doughnut-like shape and diffracts when
it propagates in free space. In quantum information they have an
enormous potential for codifying information beyond two levels using their
topological charge value~\cite{PhysRevLett.88.013601}. In other
fields, such as biophotonics for example, they are useful due to their
ability to affect the motion of particles (microorganisms) through
angular momentum transfer~\cite{nature1}. Other scientific and
technological applications for optical vortices are found in optical
systems communication, spintronics and optical tweezers~\cite{nature2,
Scheuer09071999,PhysRevLett.86.4358}. These potential applications of
optical vortices have sparkled the interest of the scientific
community on their basic properties and characteristics.

Diffraction is a fundamental phenomenon which leads to beam broadening
upon propagation. In a nonlinear medium, self-focusing reduces
diffraction whereas self-defocusing enhances the beam spreading. In a
situation where the nonlinear self-focusing effects exactly balances
diffraction, the beam can propagate as an optical spatial soliton,
i.e. a self-trapped optical beam which preserves its shape upon
propagation. Recently, spatial optical solitons have become attractive
for several technological applications. They can be defined as
self-localized solutions of nonlinear wave equations found in various
physical systems~\cite{RevModPhys.83.247}. Typical equations of this
type in optics are the nonlinear Schr\"odinger equation (NLSE) for
conservative systems, and the complex Ginzburg-Landau equation (CGLE)
for dissipative ones~\cite{nls,0963-9659-4-4-005}. The CGLE is a
master model in which dissipative solitons~\cite{akhm05} are probably
its most interesting solutions. In conservative models, such as the
ones described by the NLSE or its several variants, exchange
of energy with the surroundings is not allowed. Self-localized solutions for
the nonlinear Schr\"odinger equation originate from a balance between
nonlinearity (e.g., the Kerr effect) and dispersion/diffraction. In
contrast, for dissipative systems the solutions also exchange energy with
an external source, making the problem more complex and rich. In this case, 
an extra balance between gain and loss is required in order to obtain
stationary solutions. In particular, dissipative vortex solitons in
continuous media have been found to exist for several values of $S$,
and they are stable in wide regions of the parameter space of the
CGLE~\cite{Soto-Crespo:09,PhysRevLett.105.213901}.

In this paper we concentrate on dissipative systems governed by the
Ginzburg-Landau equation. This model appears in many branches of
science such as nonlinear optics, Bose-Einstein condensates,
chemical reactions, super-conductivity and many
others~\cite{akhm0508,RevModPhys.74.99}. Nonlinear periodic
structures offer alternative ways to control light
propagation by modifying its diffraction properties through the
modulation of its refractive index. For instance, photonic crystals
are structures of alternating refractive index that provide
unprecedented control over light fields propagating through them;
recent works show that lasers with square-lattices photonic crystal
cavities posses enhanced functionality and performance when compared
to conventional lasers~\cite{nature3}. These systems can be analyzed
in the framework of a set of coupled, linear equations which, in
solid-state physics is known as the {\it tight-binding} approximation,
while in an optics context, it is known as the coupled-mode approach.
Stationary solutions obtained in this framework are called discrete
solitons. In particular, discrete vortex solitons in conservative
systems have been reported on several theoretical and experimental
works~\cite{PhysRevA.83.063822,PhysRevE.64.026601,
PhysRevLett.92.123903,PhysRevLett.102.224102}, while 
dissipative discrete solitons have been found, analytically
and numerically, in one dimensional waveguide
arrays~\cite{SotoCrespo2003126,Maruno2005231,PhysRevA.76.043839}.

In this paper we report the finding of a wide region in the parameter
space of the discrete cubic-quintic complex Ginzburg-Landau equation
where different discrete vortex solitons coexist. 
All the individual solutions we examine in this paper posses
simultaneously two topological charges, as those reported in some of
our recent works~\cite{PhysRevA.82.063818,PhysRevA.83.043837}.
We have studied their interactions and as a result
the formation of bound states.

The rest of the paper is organized as follows. In Section \ref{mod} we
introduce the complex cubic-quintic Ginzburg-Landau model that we will
use in this work. Section \ref{res1} describes the new families of
solutions we have found, and in Section \ref{bnd}  we show 
the composite structures obtained when we let them interact
as they evolve. Section \ref{schr} analyzes the discrete nonlinear
Schr\"odinger equation case for all the solutions previously
mentioned. Finally, section \ref{con} summarizes our main results and
conclusions.

\section{Model}
\label{mod}

Optical beam propagation in nonlinear, periodical two-dimensional
waveguide array can be modeled by the following equation:
\begin{eqnarray}
\nonumber
&i\dot\psi_{m,n}+\hat C\psi_{m,n}+|\psi_{m,n}|^{2}\psi_{m,n}+\nu|\psi_{m,n}|^{4}\psi_{m,n}=\\
&i\delta\psi_{m,n}+i\varepsilon|\psi_{m,n}|^{2}\psi_{m,n}+i\mu|\psi_{m,n}|^{4}\psi_{m,n}\ .
\label{dgl2}
\end{eqnarray}
Equation(\ref{dgl2}) is the discrete version of the complex
cubic-quintic Ginzburg-Landau (CQGL) equation. Here, $\psi_{m,n}$ is the
complex field amplitude at the $(m,n)$ lattice site and
$\dot\psi_{m,n}$ denotes its first derivative with respect to the
propagation coordinate $z$. The set
\begin{displaymath}
\{m=-M,...,M\}\times\{n=-N,...,N\},
\end{displaymath}
defines the array, with $2M+1$ and $2N+1$ being the number of sites in
the horizontal and vertical directions, respectively. The {\it tight
  binding} approximation establishes that the field propagating in
each waveguide interacts linearly only with nearest-neighbor fields
through their evanescent tails. This interaction is described by the
discrete diffraction operator
\begin{equation*}
\hat C\psi_{m,n}=C(\psi_{m+1,n}+\psi_{m-1,n}+\psi_{m,n+1}+\psi_{m,n-1}),
\end{equation*}
where $C$ is a complex parameter. Its real part denotes the strength
of the coupling between adjacent sites and its imaginary part denotes
the gain or loss originated by this coupling. The nonlinear
higher-order Kerr term is represented by $\nu$ while $\varepsilon>0$
and $\mu < 0$ are the coefficients for cubic gain and quintic losses,
respectively. Linear losses are accounted by a negative value of $\delta$.

In contrast to the conservative discrete nonlinear
Schr$\ddot{\text{o}}$dinger (DNLS) equation, the optical power,
defined as
\begin{equation}
Q(z)=\sum_{m,n=-M,-N}^{M,N}|\psi_{m,n}(z)|^2\ ,
\end{equation}
is not a conserved quantity in the present model. Nevertheless, for a
self-localized solution, the power and its evolution will be the main
quantity that we will monitor in order to identify different families
of stationary solutions.

We look for stationary solutions of Eq.(\ref{dgl2}) of the form
$\psi_{m,n}(z)=\phi_{m,n}\exp(i\lambda z)$ where $\lambda$ is real and
$\phi_{m,n}$ are complex amplitudes. We are interested in
solutions localized in space whose phase changes azimuthally by an
integer number ($S$) of $2\pi$ along a discrete closed-circuit. In
such a case, the self-localized solution is called a discrete vortex
soliton~\cite{Pelinovsky200520} with vorticity $S$. By inserting the
previous {\em ansatz} into Eq. (\ref{dgl2}) we obtain the following
set of $(2M+1)\times(2N+1)$ algebraic coupled complex equations:
\begin{eqnarray}
\nonumber
&-\lambda\phi_{m,n}+\hat C\phi_{m,n}+|\phi_{m,n}|^{2}\phi_{m,n}+\nu|\phi_{m,n}|^{4}\phi_{m,n}=\\
&i\delta\phi_{m,n}+i\varepsilon|\phi_{m,n}|^{2}\phi_{m,n}+i\mu|\phi_{m,n}|^{4}\phi_{m,n}\ .
\label{adgl2}
\end{eqnarray}
We solve Eq.(\ref{adgl2}) by using a multi-dimensional
Newton-Raphson iterative algorithm. The method requires an initial
guess, and it converges rapidly when using a highly-localized profile
[more details can be found in Ref.~\cite{PhysRevA.82.063818}].

\subsection{Linear stability analysis}
\label{sta}
Small perturbations around the stationary solution can grow
exponentially, leading to the destruction of the vortex soliton. A
stability analysis provides us with the means for establishing which
solutions are linearly stable. Let us introduce a small perturbation
$\tilde\phi$, to the localized stationary solution
\begin{equation}
\psi_{m,n}=[\phi_{m,n}+\tilde\phi_{m,n}(z)]e^{i\lambda z},\hspace{1cm}\tilde\phi_{m,n}\in\mathbb{C}
\label{Eq:pert},
\end{equation}
then, after replacing Eq.(\ref{Eq:pert}) into Eq.(\ref{dgl2}) and then
linearizing with respect to $\tilde\phi$, we obtain:
\begin{eqnarray}
\nonumber
&\dot{\tilde\phi}_{m,n}+\hat C\tilde\phi_{m,n}-i\delta\tilde\phi_{m,n}+\\
\nonumber
& [2(1-\varepsilon)|\phi_{m,n}|^{2}+3(\nu-\mu)|\phi_{m,n}|^{4}-\lambda]\tilde\phi_{m,n}+\\
&[(1-\varepsilon)\phi_{m,n}^{2}+2(\nu-\mu)|\phi_{m,n}|^{2}\phi_{m,n}^{2}]\tilde\phi^{*}_{m,n}=0.\
\label{dgl2p}
\end{eqnarray}
The solutions of the above homogeneous linear system can be written as
\begin{equation}
\tilde\phi_{m,n}(z)=C^{1}_{m,n}\exp{[\gamma_{m,n}z]}+C^{2}_{m,n}\exp{[\gamma_{m,n}^{*} z]}
\label{solpert},
\end{equation}
where $C^{1,2}$ are integration constants and $\gamma_{m,n}$ is the
discrete spectrum of the $eigensystem$ associated with (\ref{dgl2p}).
The solutions are unstable if at least one eigenvalue has a positive
real part, that is, if $\textsf{max}\{\text{Re}(\gamma_{m,n})\}>0$.

\section{Multiplicity of stable vortex soliton families}
\label{res1}

\begin{figure}
\centering
\includegraphics[width=0.75\linewidth]{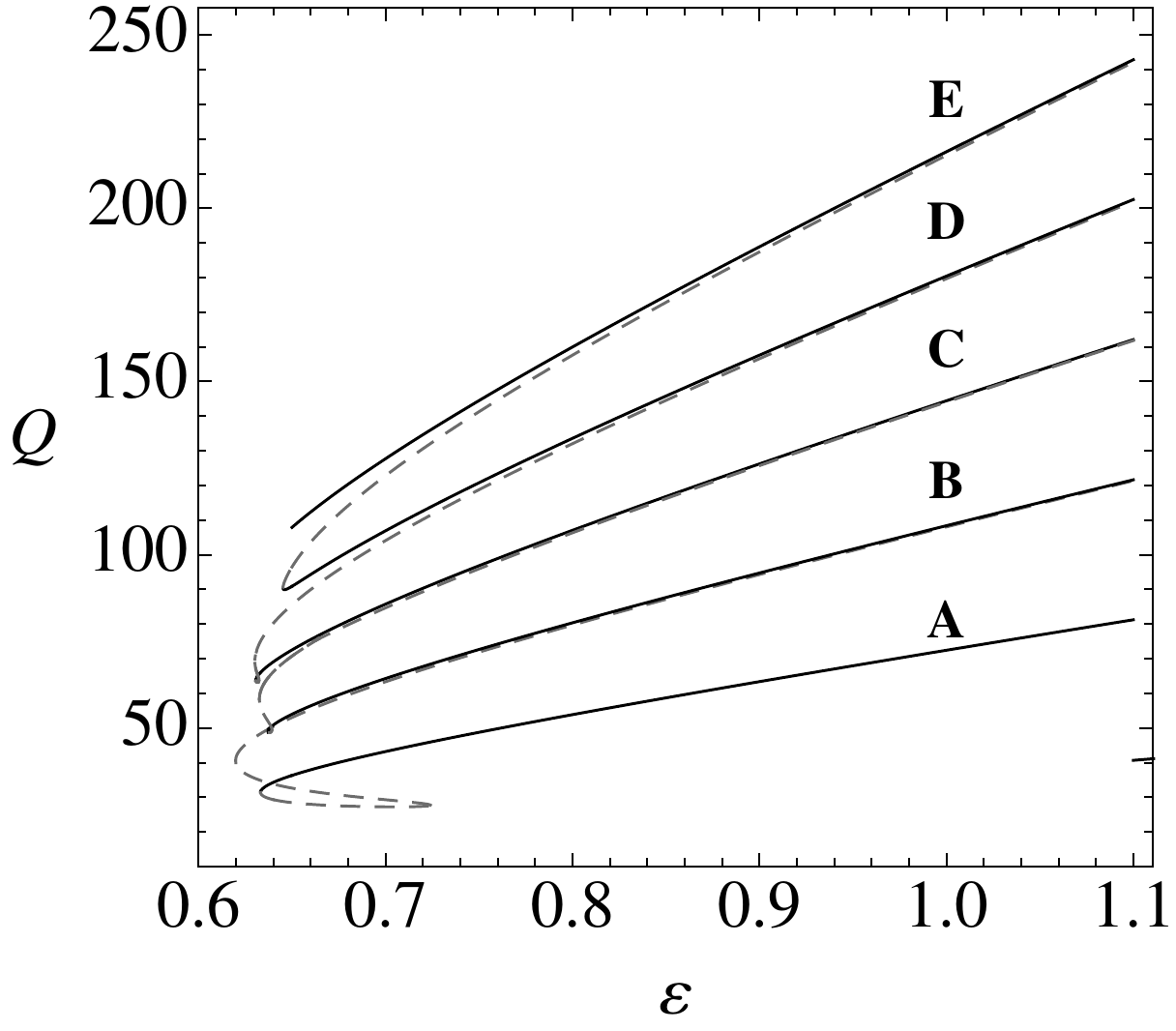}
\caption{$Q$ versus $\varepsilon$ diagram for several vortex interconnected
soliton families. Solid lines correspond to stable families while dashed lines
to unstable ones. (CQGL equation parameters: $C=0.8$, $\delta=-0.9$, $\mu=-0.1$,
$\nu=0.1$).}
\label{fig1}
\end{figure}

\begin{figure}[ht]
\centering
\begin{tabular}{ccccc}
A&B&C&D&E\\
\includegraphics[width=0.18\linewidth]{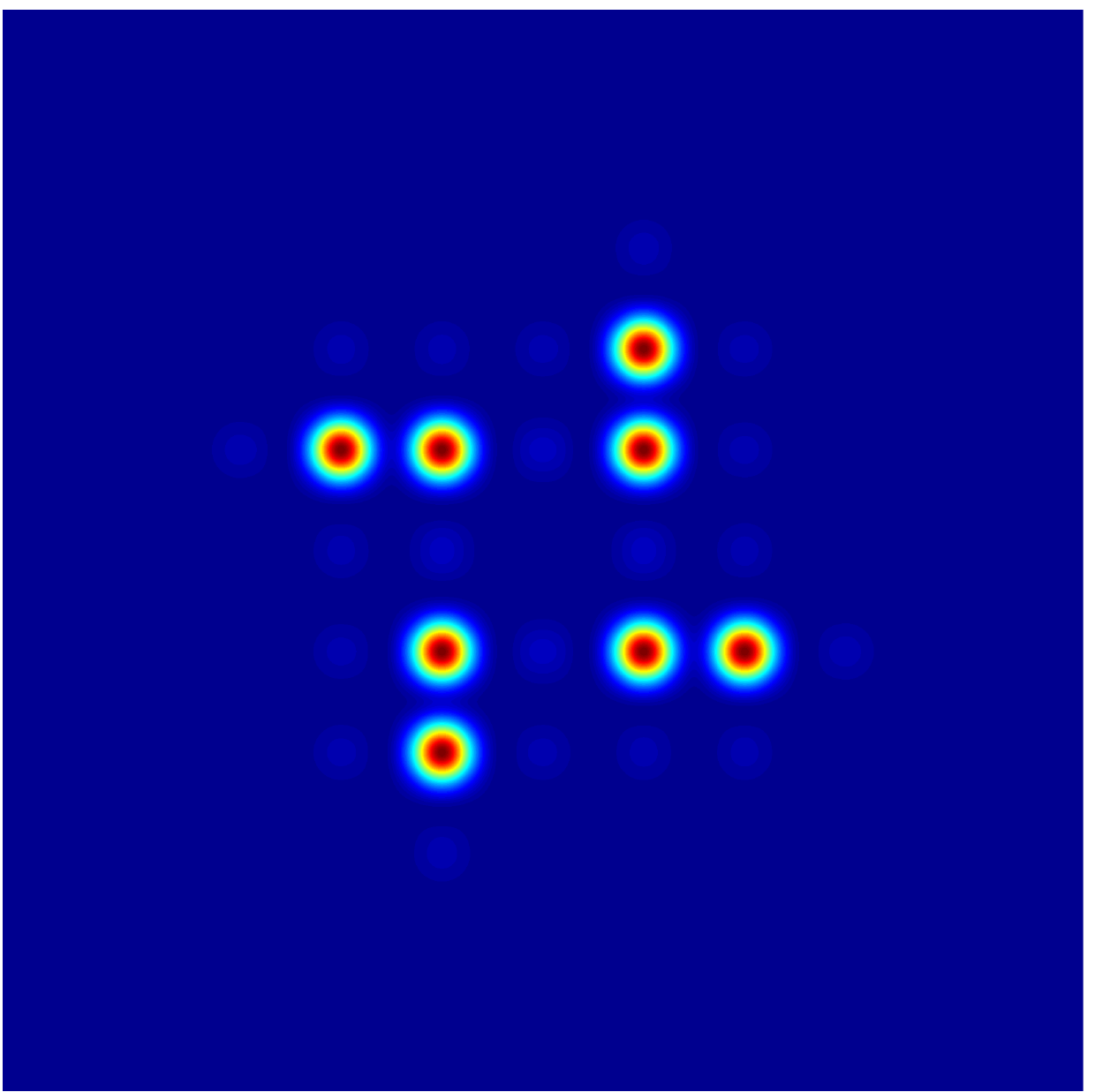}&

\includegraphics[width=0.18\linewidth]{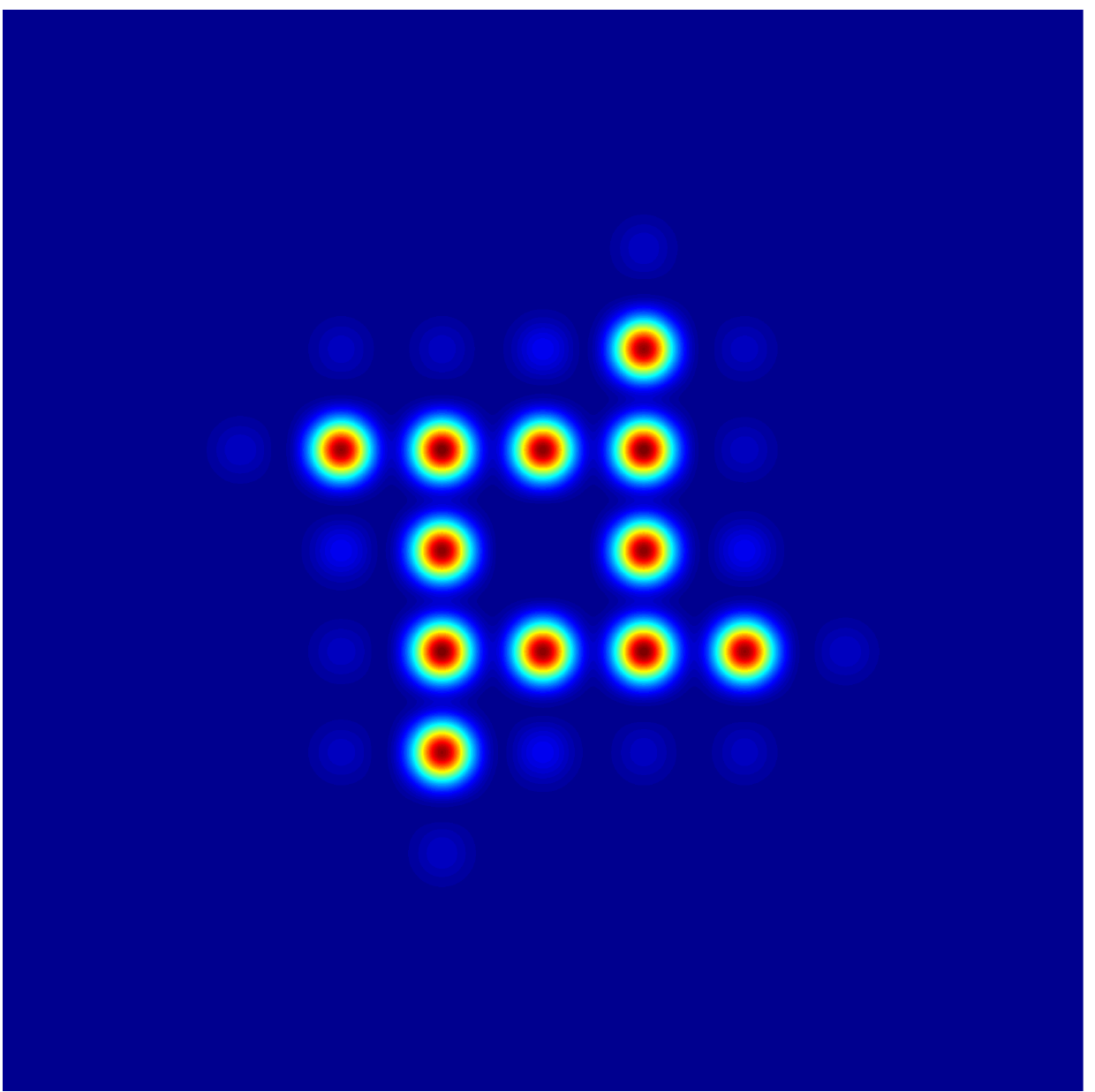}&

\includegraphics[width=0.18\linewidth]{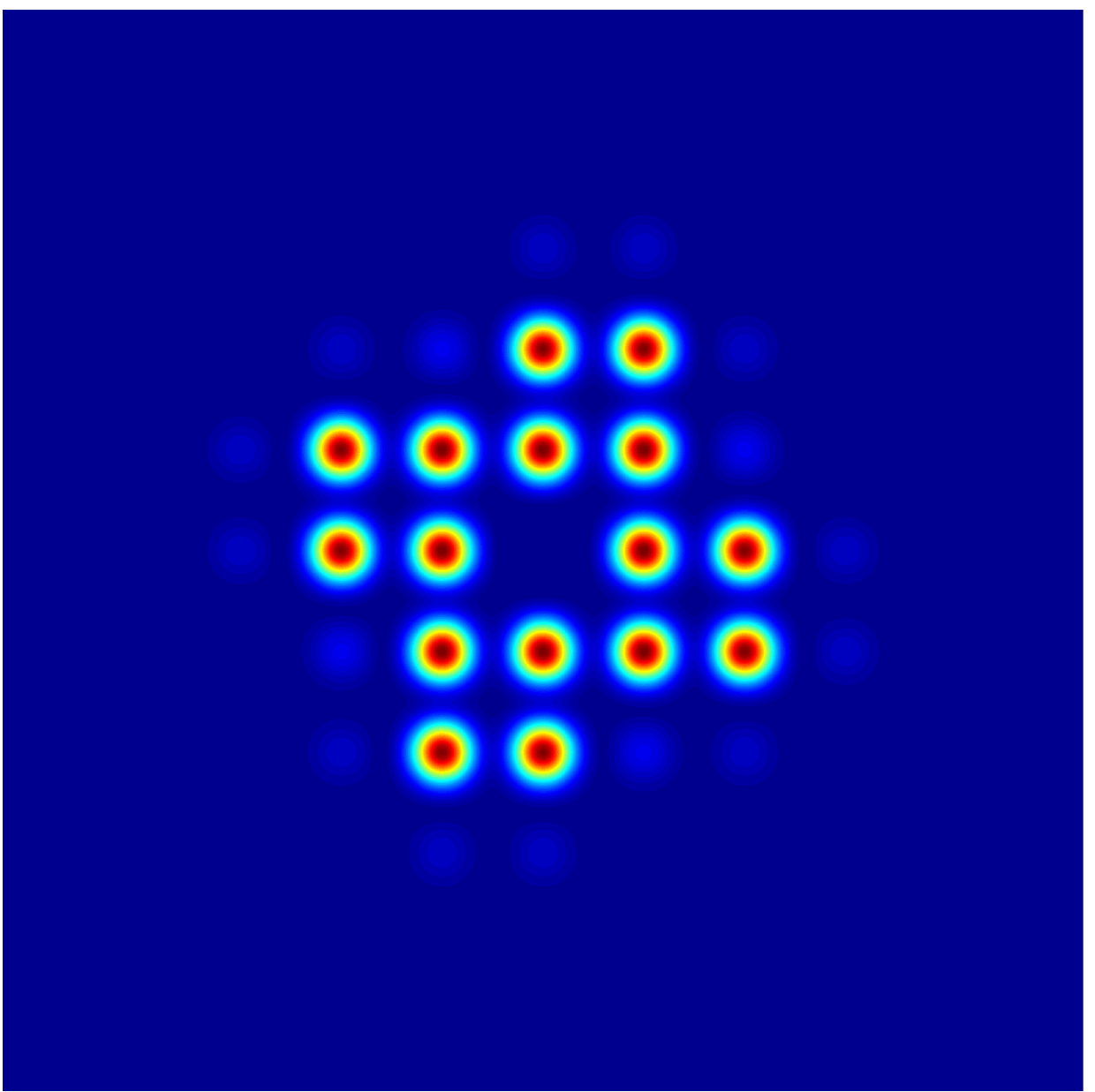}&

\includegraphics[width=0.18\linewidth]{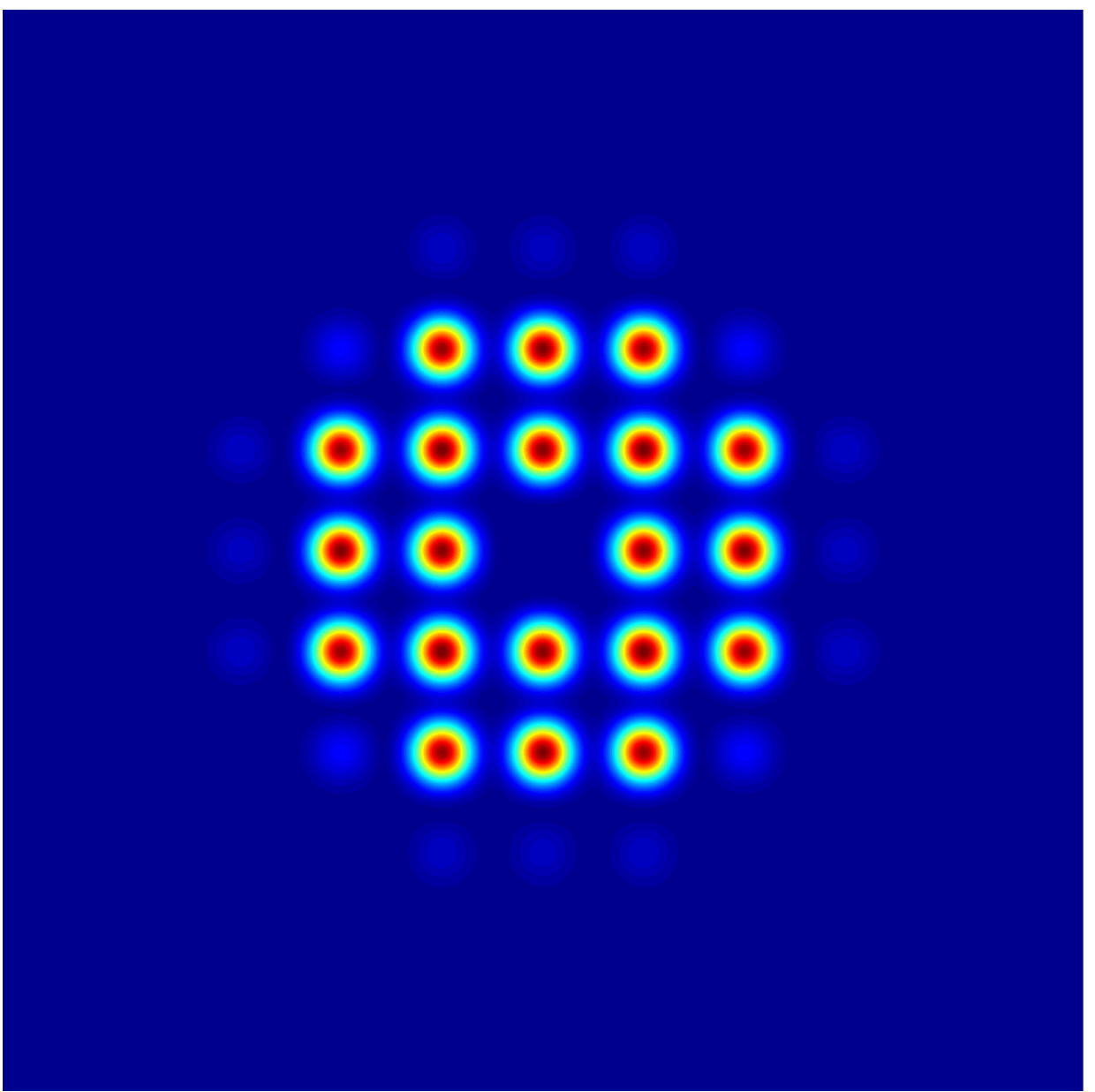}&

\includegraphics[width=0.18\linewidth]{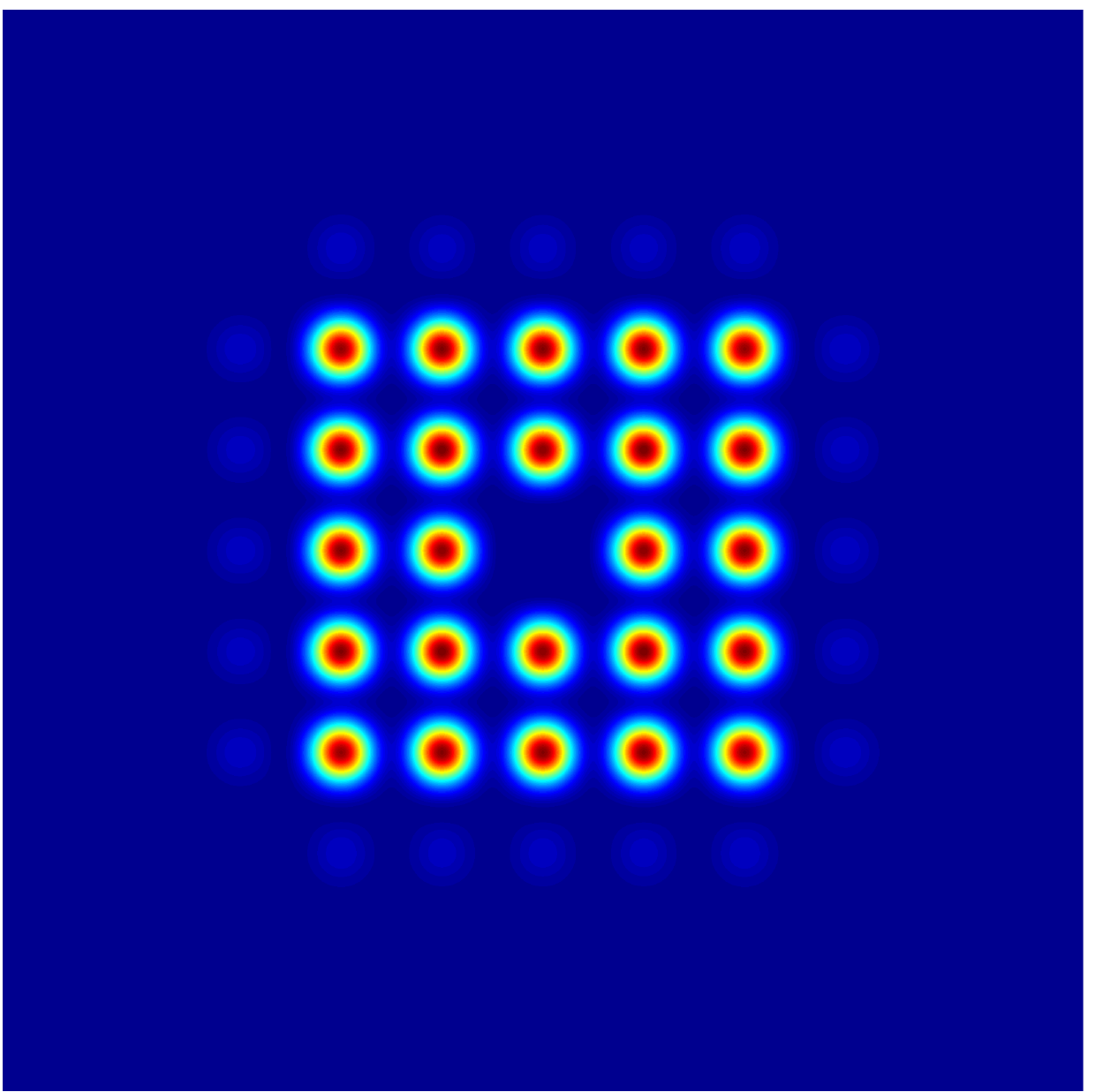}\\

\includegraphics[width=0.18\linewidth]{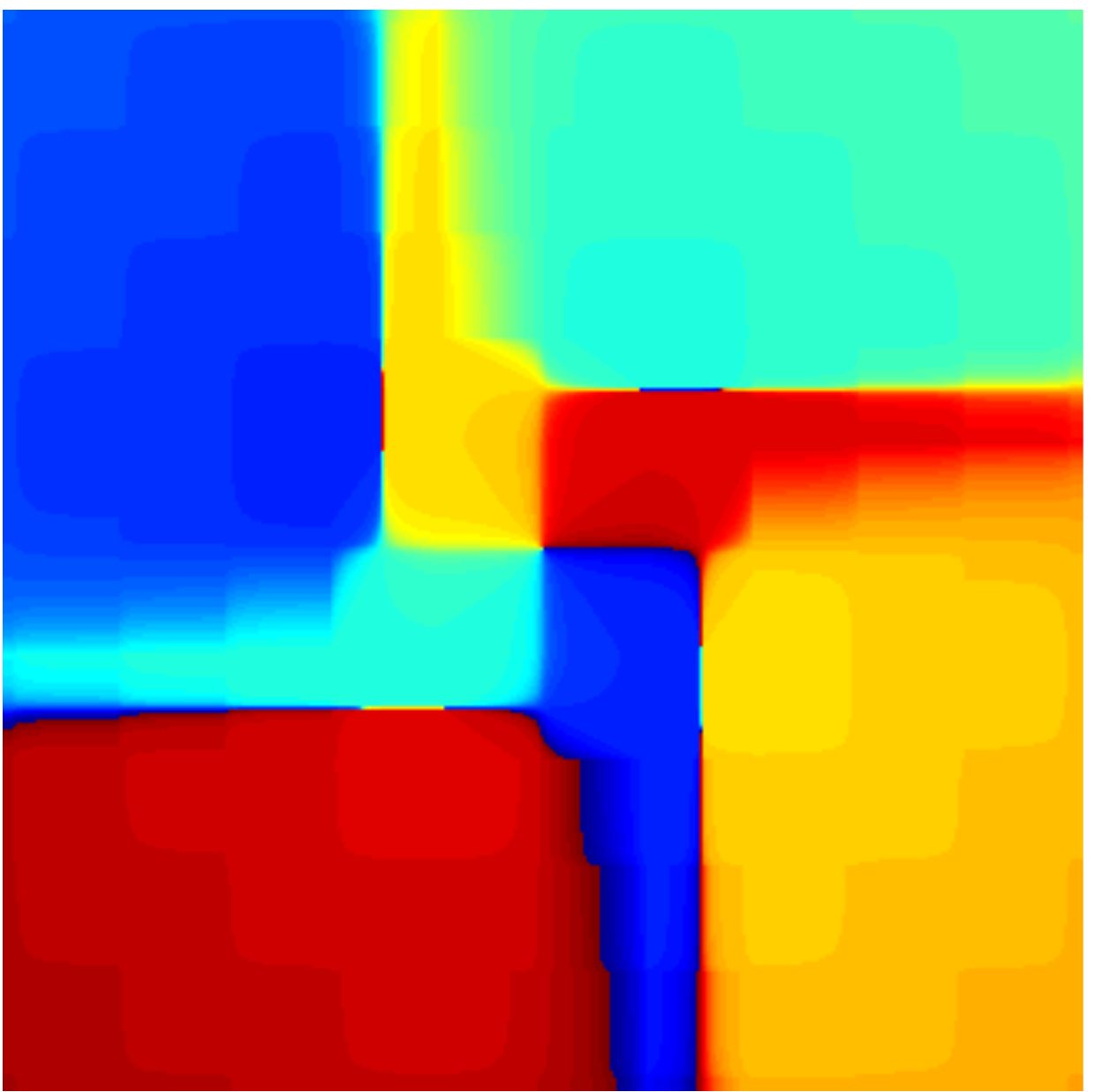}&

\includegraphics[width=0.18\linewidth]{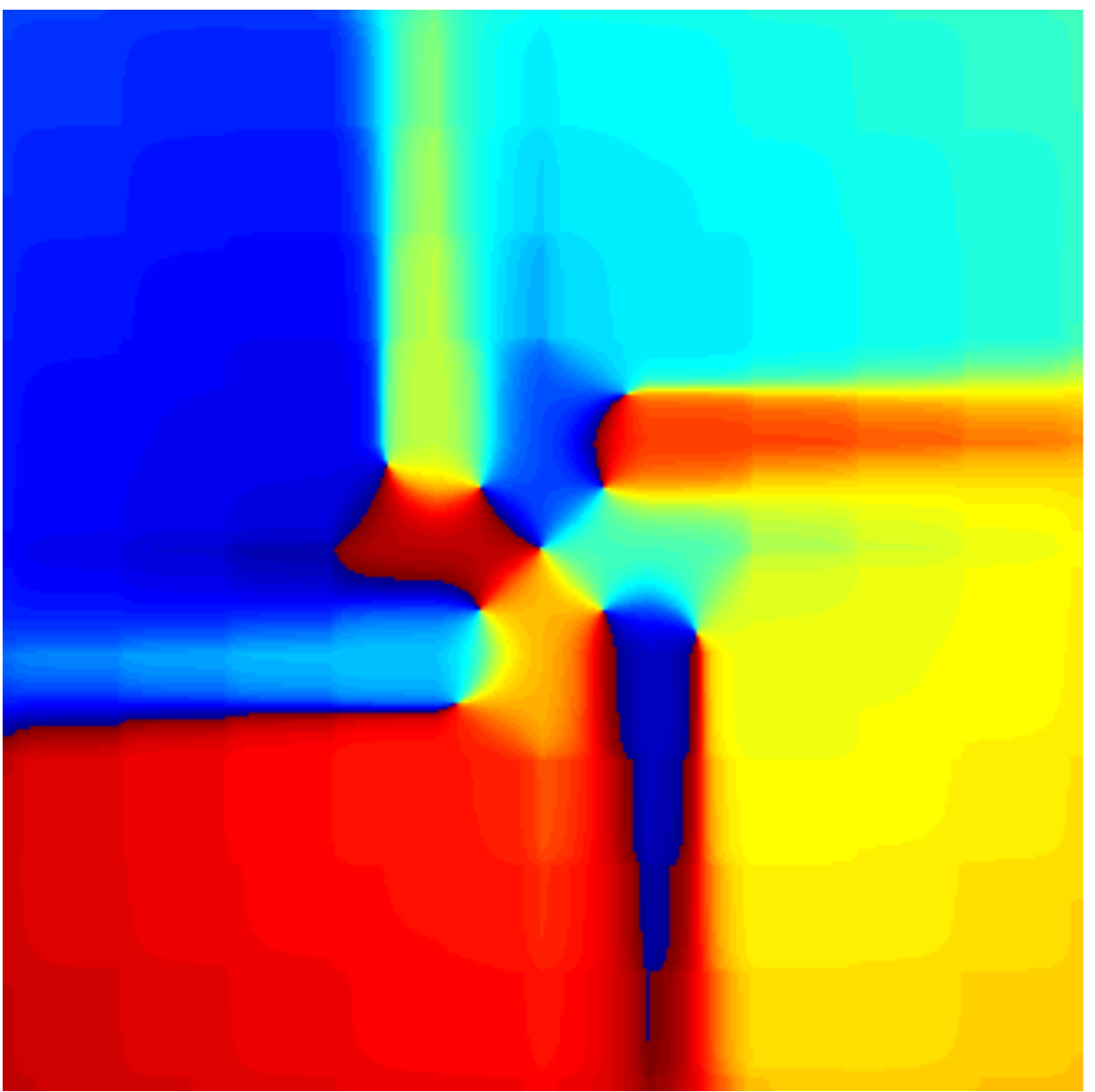}&

\includegraphics[width=0.18\linewidth]{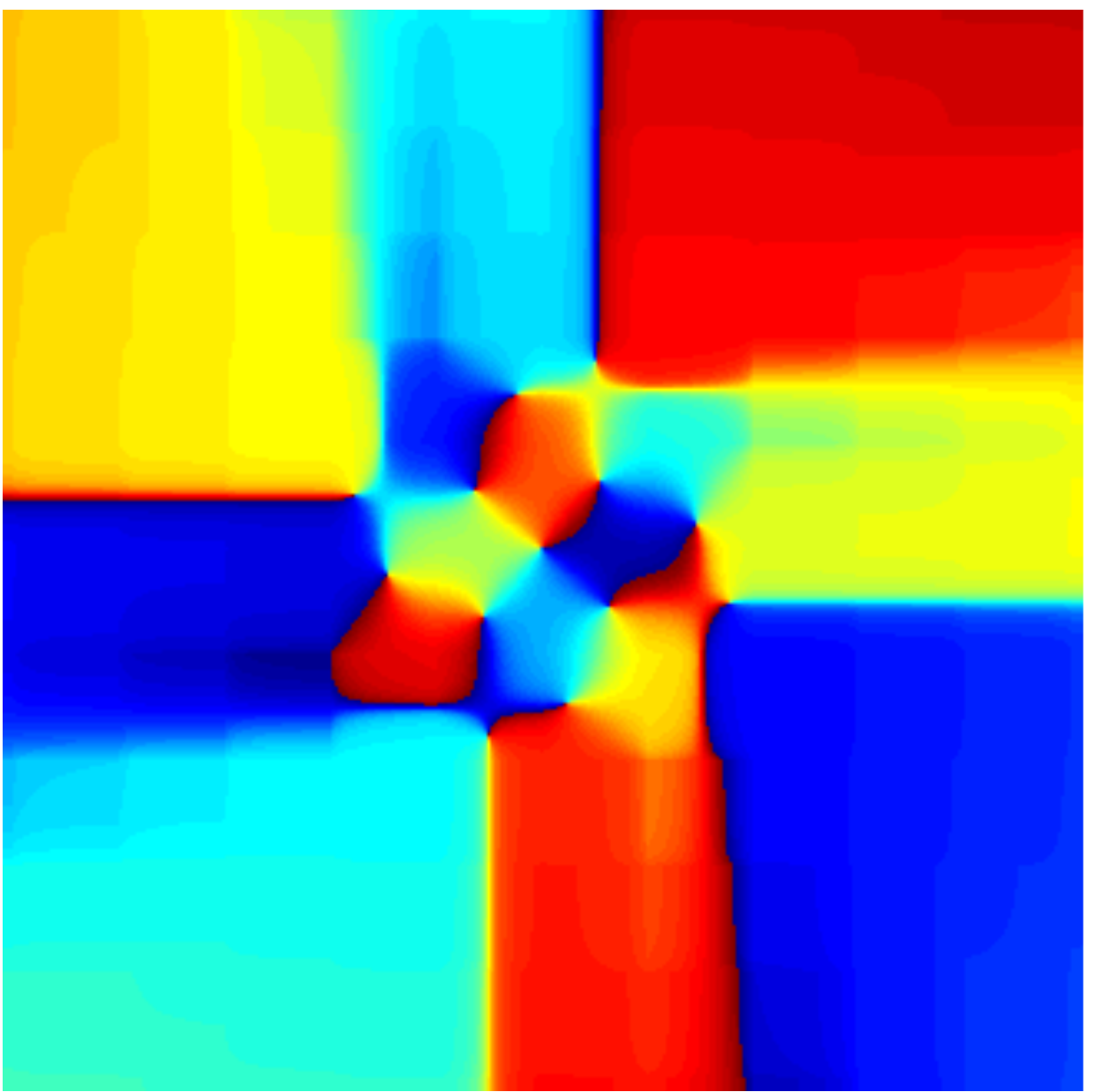}&

\includegraphics[width=0.18\linewidth]{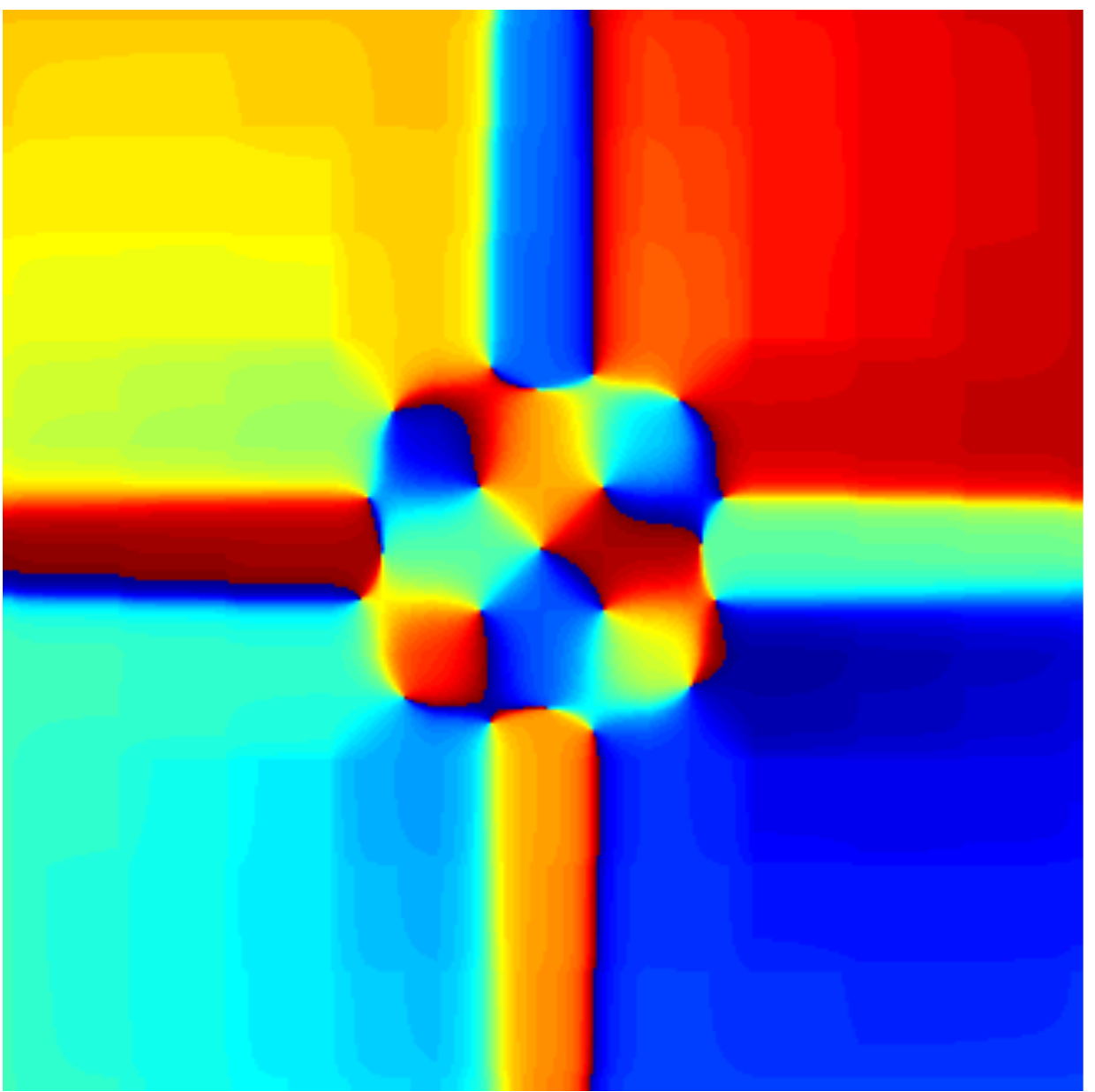}&

\includegraphics[width=0.18\linewidth]{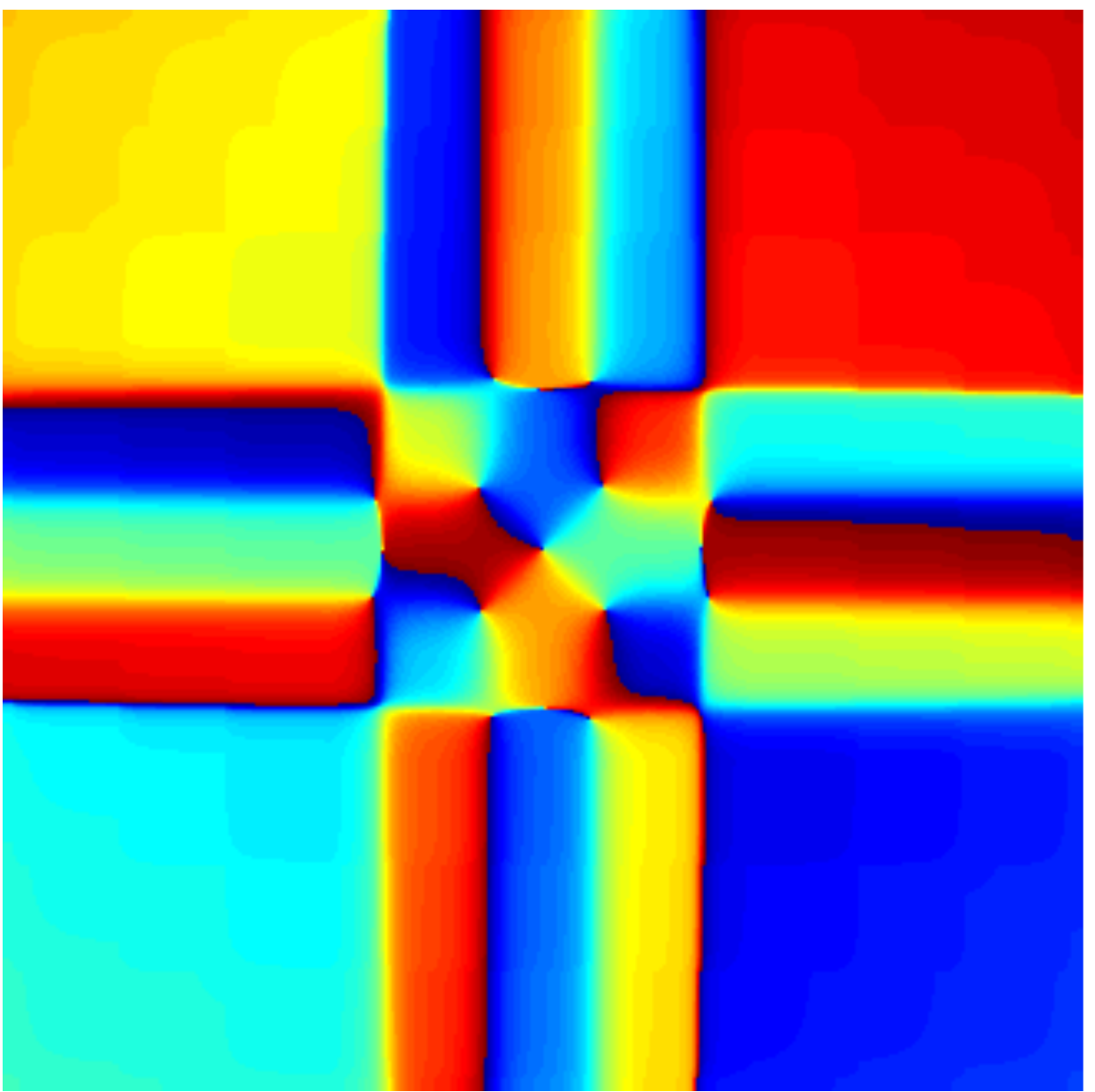}

\end{tabular}
\caption{(Color online) Color map plots for the amplitude (top) and phase
(bottom) profiles of stable vortex solutions for the families marked in
Fig.(\ref{fig1}).}
\label{fig2}
\end{figure}

As stated above, the nonlinear gain in the system is mainly controlled
by $\varepsilon$; this parameter will be the only one that we will
change in our simulations. Once we find a stationary solution - for a
fixed set of parameters -, we compute its linear stability, and 
then change the parameters slightly and find the 
new solution using the previous one as an {\em ansatz}. In this manner
we obtain all the solution families displayed in the $Q$ 
vs $\varepsilon$ diagram shown in Fig.(\ref{fig1}). 
The first of them ({\bf A}-family), was already
reported in our recent
works~\cite{PhysRevA.82.063818,PhysRevA.83.043837}. It was obtained by
starting from the fundamental four-peaks discrete vortex soliton with
$S=1$, after passing throughout several saddle-node points. A striking
property of all the solutions shown in Fig.(\ref{fig1}) is that they
have - simultaneously - two topological charges; i.e., they are
two-charge vortices. Representative amplitude and phase profiles for
these families are shown in Fig.(\ref{fig2}).

Stable families {\bf B}, {\bf C}, {\bf D} and {\bf E} - shown in
Figs.(\ref{fig1}) and (\ref{fig2}) - were unveiled after observing the
dynamical evolution of solutions belonging to unstable branches
(dashed gray lines). In some cases,
the stable and unstable families are so close that they are
nearly indistinguishable at the scale of Fig.(\ref{fig1}). From the
amplitude profiles we can see that there is a difference of four
excited sites between one stable family and the next one. Family
{\bf A} has eight main excited sites and family {\bf E} has twenty
four main peaks. All these families show a coexistence of two
topological charges.

The amplitude profile for case ({\bf A}) shows a swirl spatial
  configuration. From its phase profile we can identify a topological
  charge $S=1$ in the core - the most inner discrete contour - and a
  charge $S=-3$ away from the center. The phase profiles for families
  {\bf B}, {\bf C}, {\bf D} and {\bf E} show a topological charge
  $S=-3$ in the core of these solutions. From {\bf B} to {\bf D}, the
  topological charge has the same value in the core and away from there, but
  the phase profile outside looks rotated respect to the center. For
  the last family, {\bf E}, the topological charge has increased up to
  $S=-7$ away from the center.

\section{Composite structures}
\label{bnd}

Next, we study the formation of bound states composed of two vortex
solitons belonging to the family {\bf E} with $\varepsilon=0.9$. 
We have chosen this family due to his high value of vorticity and
squared symmetry equivalent to that of the optical lattice. 
We study dynamically the evolution of an array of two of these solutions,
horizontally shifted by a certain small distance. We
tested two initial configurations differing in their initial
separation and, for each one of them we try a broad range of initial
phase differences, following a procedure similar to the one implemented in
Ref.~\cite{PhysRevLett.79.4047}.
For this purpose, we multiply the second solution by a phase factor $e^{i\theta_{\alpha}}$, 
 where $\theta_{\alpha}\equiv\frac{\pi}{20}\alpha$ with $\alpha=1,2,...,40$.
A bound state is reached when the relative phase 
 ($\Delta\theta$, defined as the phase difference between two given sites
in each solution) becomes a constant.  In continuous and homogeneous
systems, the separation distance also changes along the evolution and it
becomes a constant when the bound state is formed. Here, in the dissipative
discrete case we do not observe  any soliton mobility and, therefore,
the separation distance remains invariant. 

\begin{figure}[ht]
\centering
\begin{tabular}{cccc}
\includegraphics[width=0.46\linewidth]{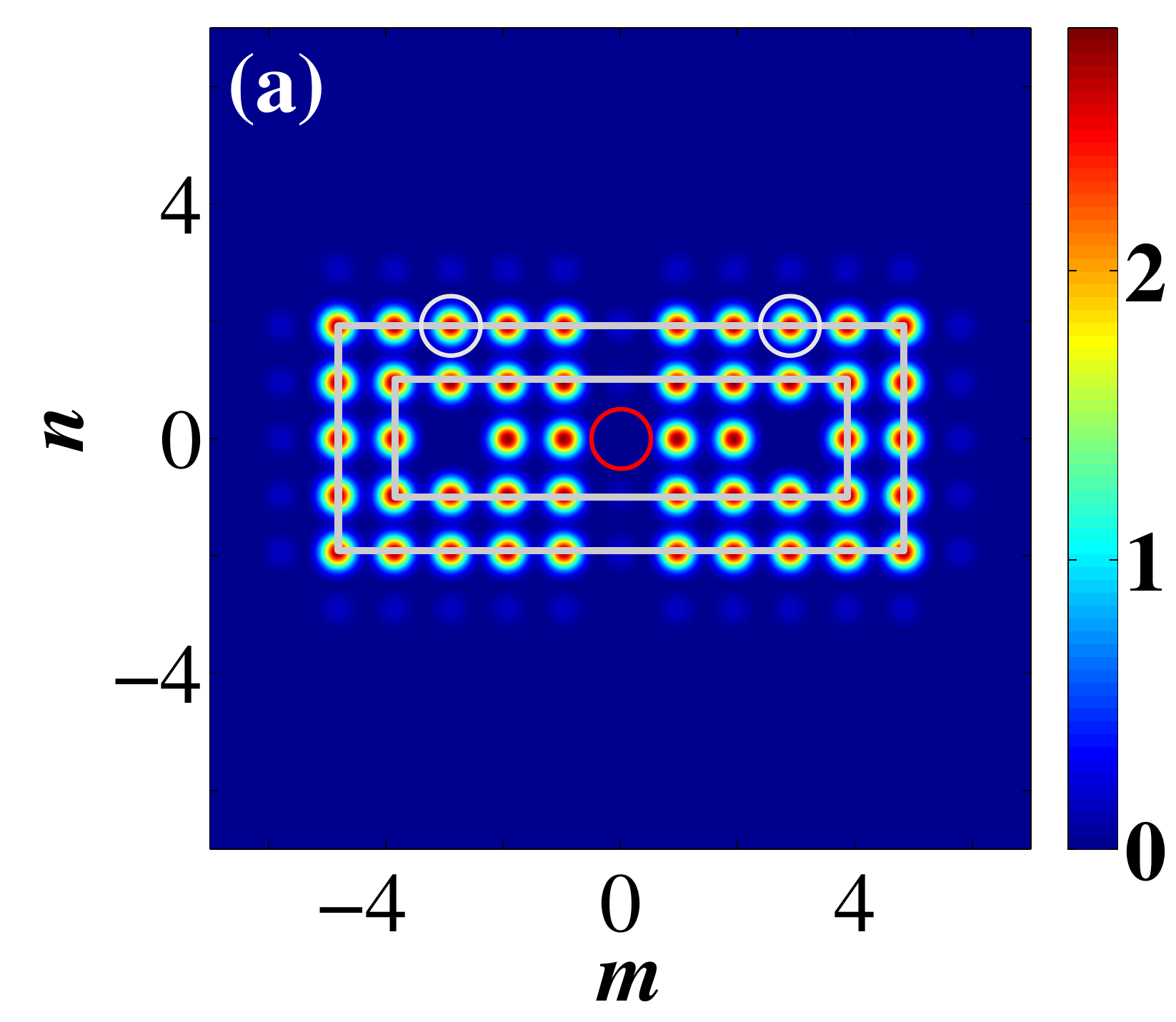}&
\includegraphics[width=0.475\linewidth]{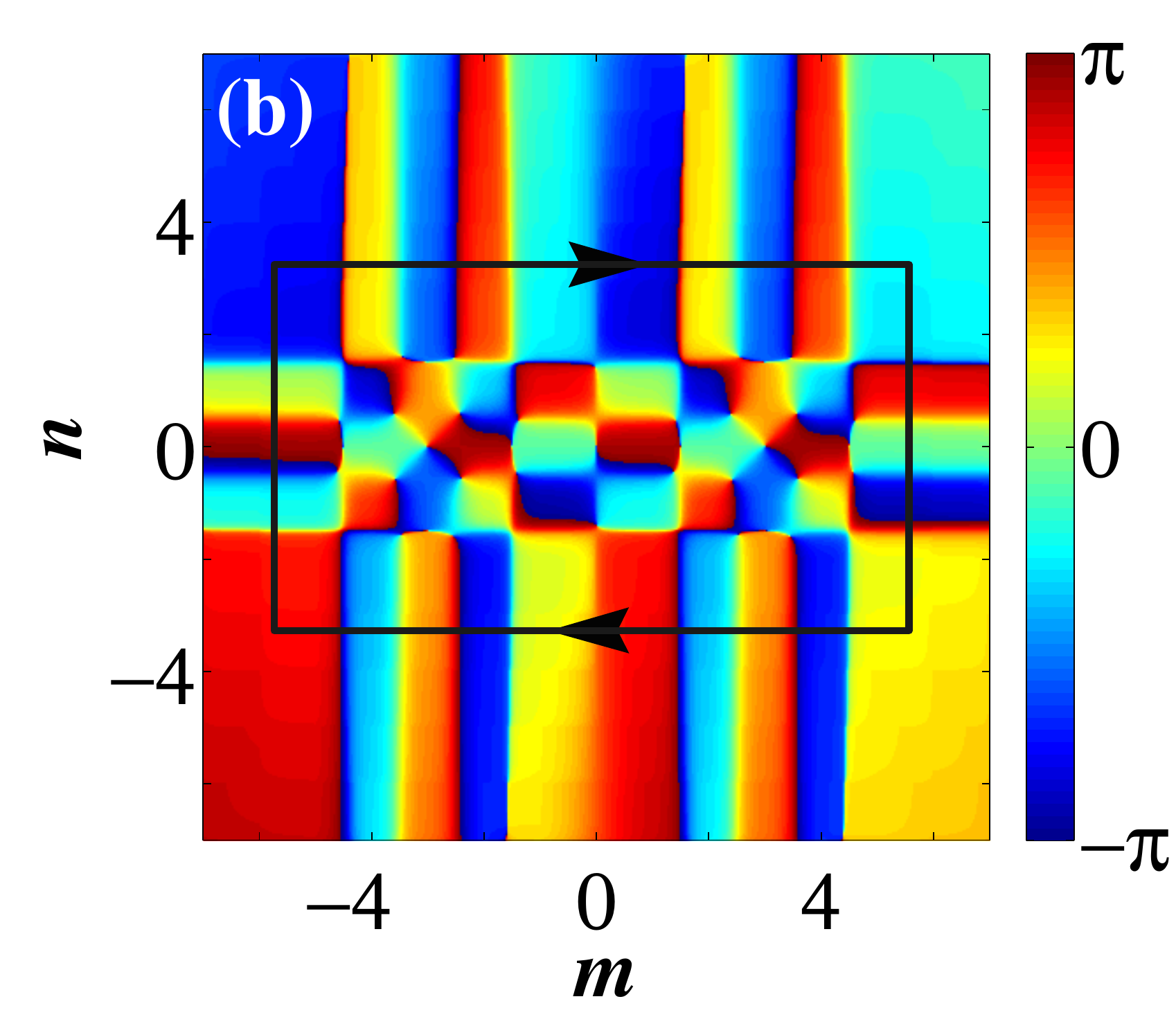}\\
\includegraphics[width=0.46\linewidth]{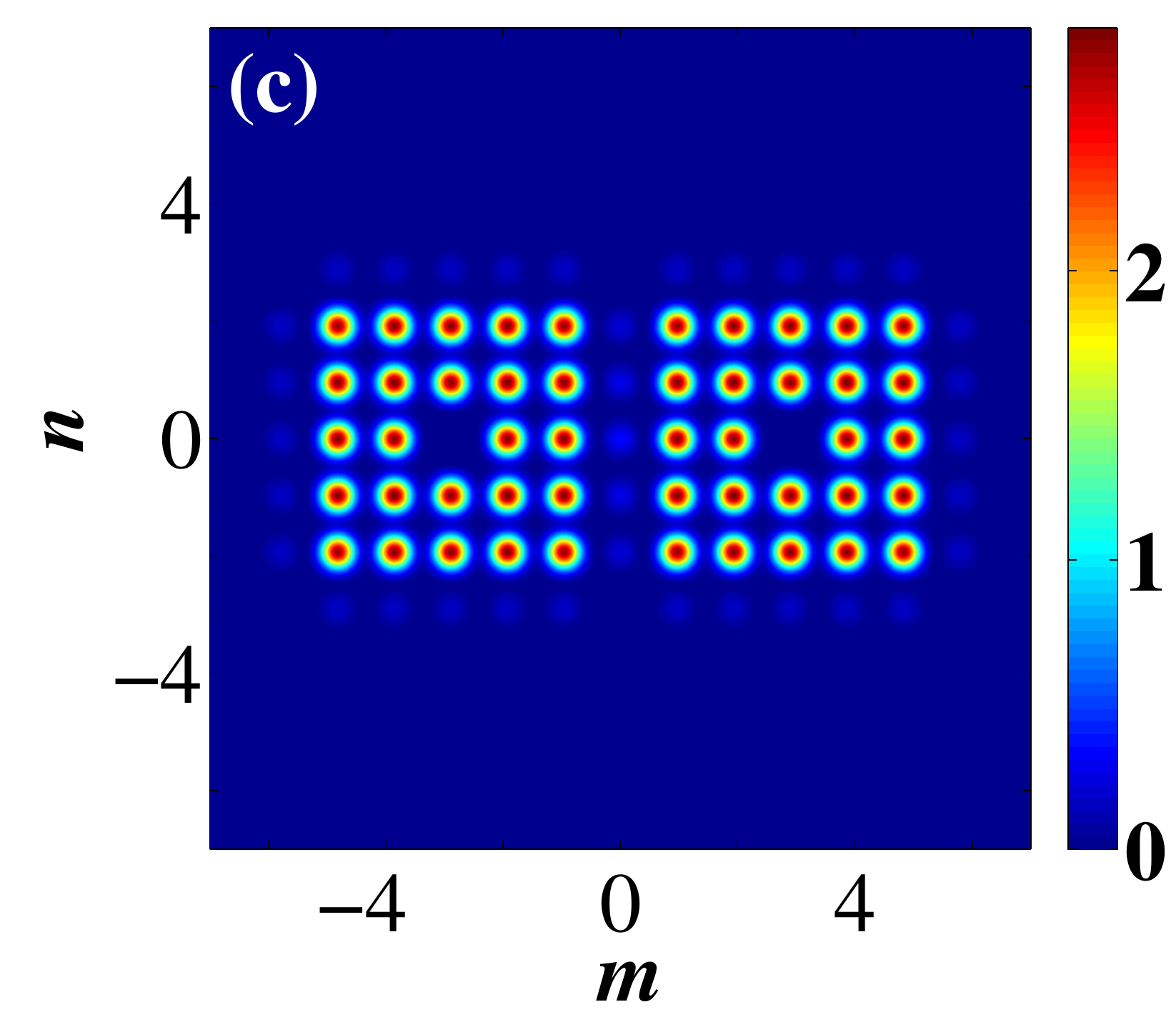}&
\includegraphics[width=0.475\linewidth]{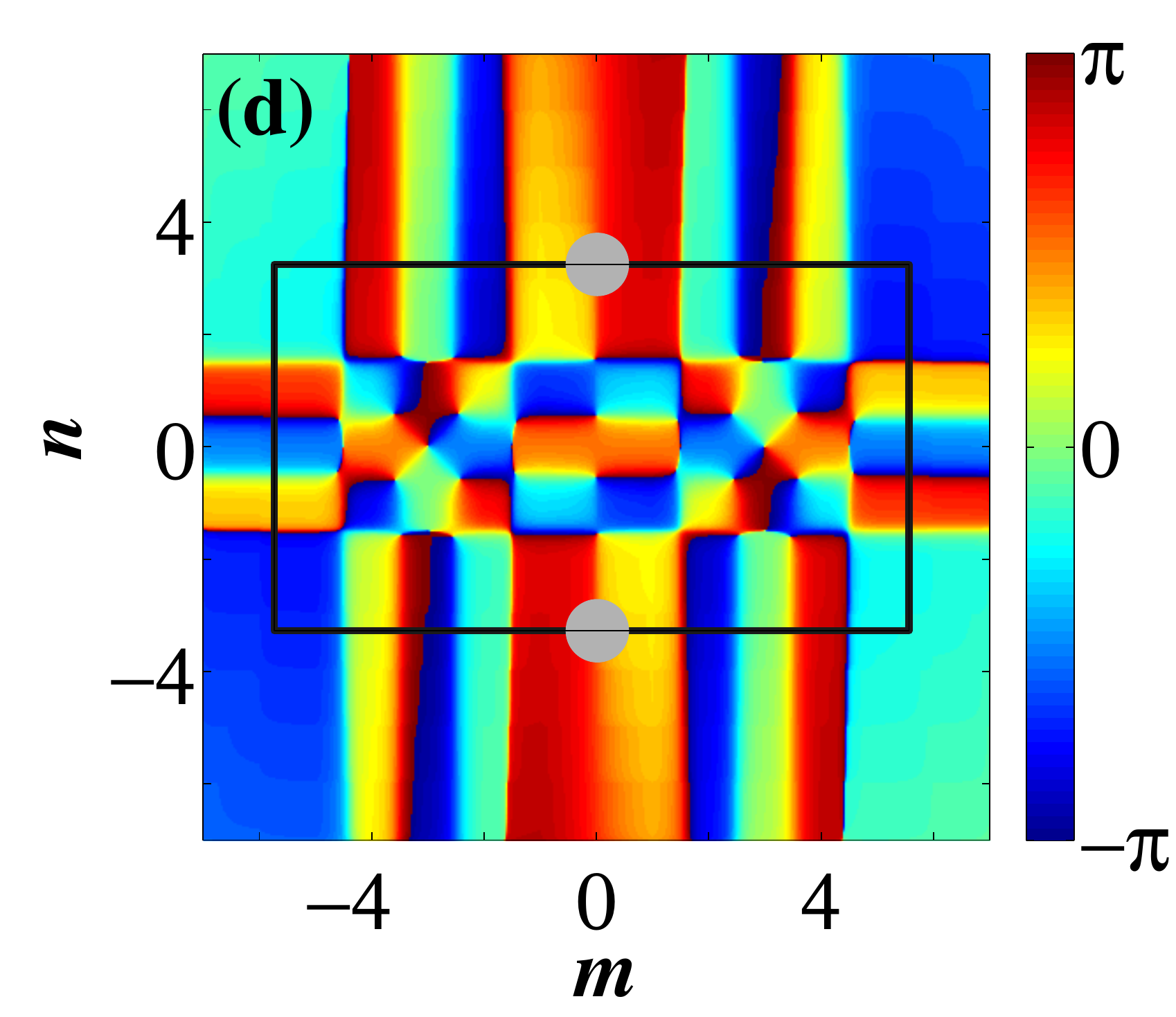}
\end{tabular}
\caption{(Color online) Color map plots showing the amplitude (left) and
phase (right) profiles for the stable solutions corresponding with the basins
of attraction shown in Fig.\ref{fig4}. For basins {\bf b1$^{\leftrightarrow}$} the
stable vortex soliton is similar to the profiles shown in (a) and (b). Profiles
for the {\bf b2} basin looks slightly different and are shown in (c) and (d).}
\label{fig3}
\end{figure}

\begin{figure}[ht]
\flushleft
\begin{tabular}{cc}
\includegraphics[width=0.46\linewidth]{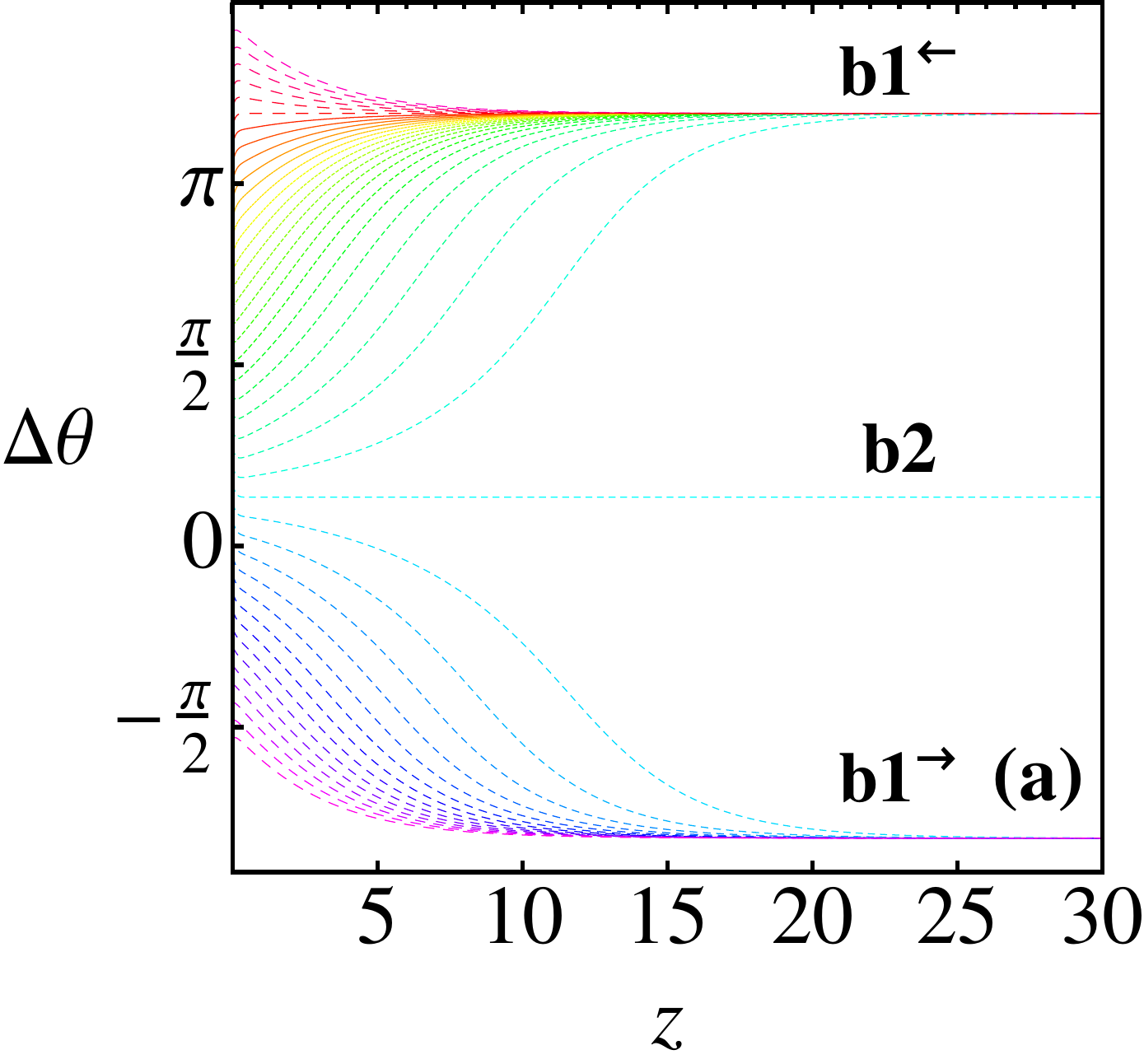}&
\includegraphics[width=0.51\linewidth]{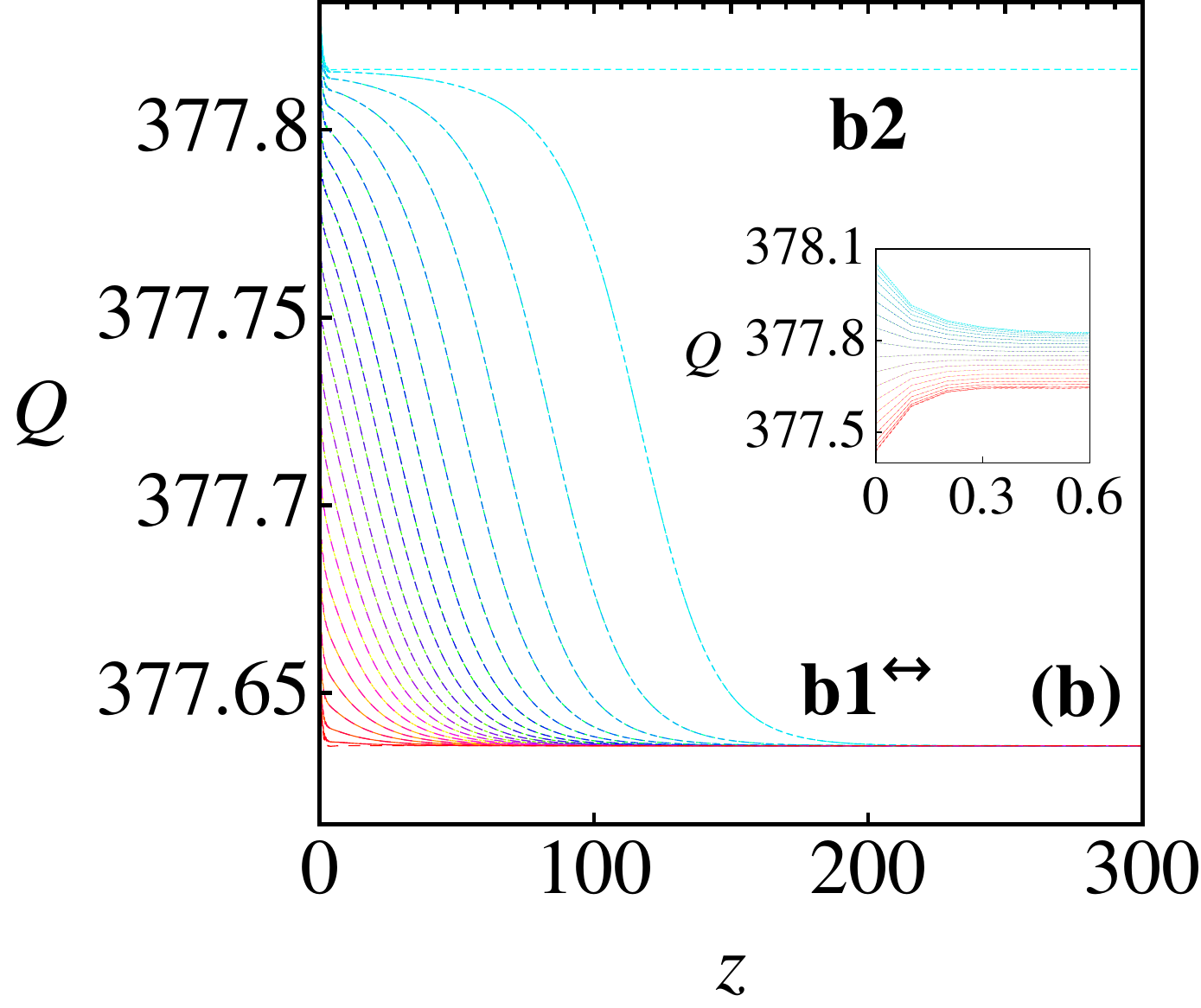}
\end{tabular}
\caption{(Color online) (a) Dynamic evolution of the relative phase between
the two sites
enclosed by the white circles in the vortex solutions shown in Fig.(\ref{fig3}).
(b) Optical power evolution for the same vortex solutions. The inset in (b) shows a
magnification of the initial stage of evolution.}
\label{fig4}
\end{figure}

We have measured the phase difference, in both configurations,
  for the sites enclosed by the white circles showed in
  Fig.\ref{fig3}(a). Figs.\ref{fig4}(a) and (b) show $\Delta\theta$
and $Q$ versus $z$, respectively, for the first configuration shown in
Fig.\ref{fig3}(a). We clearly identify two attraction basins, labeled
as {\bf b1$^{\leftarrow}$} and {\bf b1$^{\rightarrow}$}. Both ({\bf b1$^{\leftrightarrow}$})
correspond to the lower power value
shown in Fig.\ref{fig4}(b). This implies that both basins are
symmetrically equivalent solutions. The unstable configuration is
labeled as {\bf b2} and it corresponds to the upper power value in
Fig.\ref{fig4}(b) [Figs.\ref{fig3}(c) and (d) show the amplitude and
phase profiles of this unstable solution].

All these solutions preserve the central core structure, keeping the
same topological charge as the initial input condition.
Figs.\ref{fig3}(a) and (c) show the amplitude profile for both of
them; although they are very similar, the first one has an extra
central core (marked by a red circle), located at the center of the
structure. For the second structure we can note that the column
in the middle ($m=0$) is filled by small tails, without a central core.
By taking a closed look at the rectangular contour sketched in
Fig.\ref{fig3}(b), we find that the phase varies continuously. The
charge increases in the direction indicated by the arrows in this
contour, with an accumulated charge of $S=11$. On the other hand, if
we look how the phase changes along the rectangle sketched in
Fig.\ref{fig3}(d), we see that the topological charge is not well
defined on this contour. Indeed, the topological charge is truncated
(see gray filled circles) meaning that this structure is not a
composed vortex beam. Nevertheless, this profile can be thought as two
non-interacting vortex solitons with a $\pi$ radians rotation between
them. As we can see from Fig.\ref{fig4}(a), any small
  variation in the phase leads to this bound solution to evolve
  towards the basin of attraction {\bf b1$^{\leftrightarrow}$}. No 
  other initial condition goes to {\bf b2}, meaning that this
  is not properly a basin of attraction. So, we can say that 
  vorticity is a necessary condition, achieved during the propagation,
  for the stability of this kind of structures.

\begin{figure}[ht]
\center
\begin{tabular}{cc}
\includegraphics[width=0.6\linewidth]{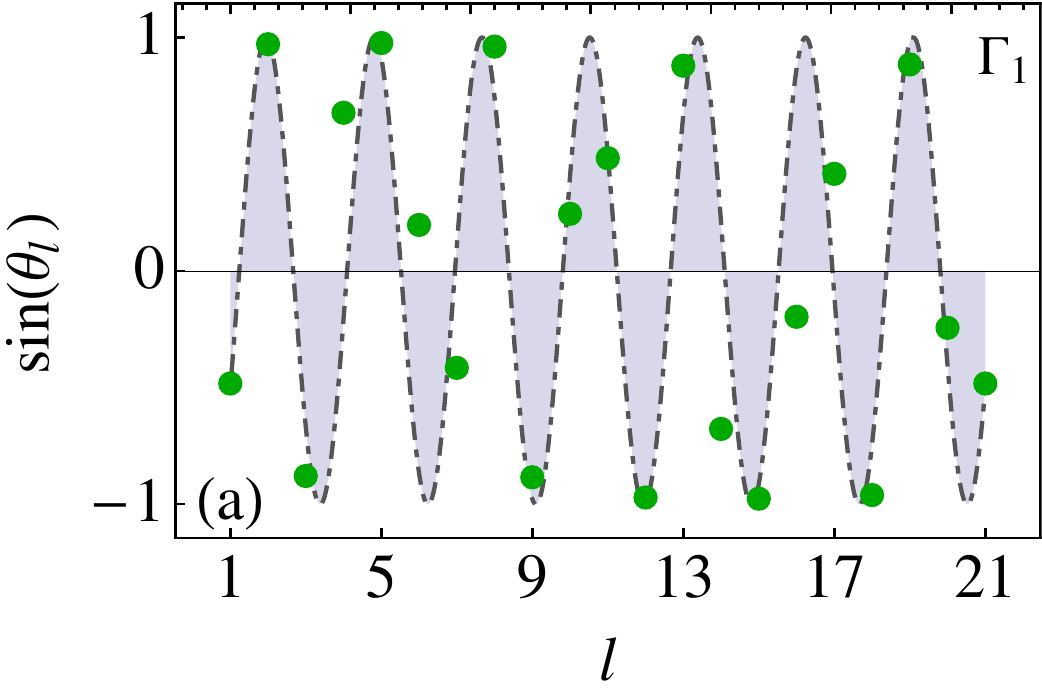}\\
\includegraphics[width=0.6\linewidth]{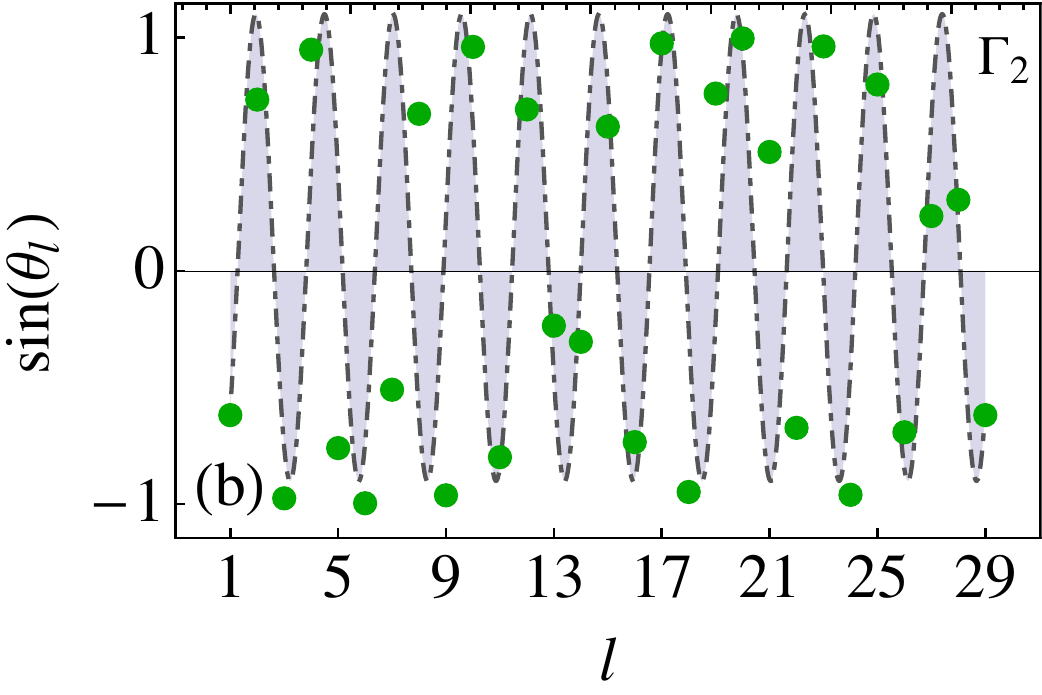}
\end{tabular}
\caption{(Color online) $\sin(\theta_{l})$ versus $l$ for contours (a) $\Gamma_{1}$ and (b) $\Gamma_{2}$}.
\label{fig5}
\end{figure}
 
For the sake of clarity, we plot $\sin(\theta_{l})$ vs $l$, where $l$
corresponds to the site on the inner ($\Gamma_{1}$) and outer
($\Gamma_{2}$) discrete contours sketched in Fig.\ref{fig3}(a).
Fig.\ref{fig5}(a) shows a good correspondence between the data (green
points) and a sinusoidal function (gray line) with seven periods ($S =
7$) along the twenty one sites of the $\Gamma_{1}$ contour. For the
$\Gamma_{2}$ contour, which contains twenty nine sites, we count
eleven periods ($S=11$) as shown in Fig.\ref{fig5}(b). Fig.\ref{fig5}
explicitly shows the different topological charges contained
simultaneously in this solution. This supports the right identification of
discrete vortex solitons, which is not an easy task for 
discrete systems.

\begin{figure}[ht]
\begin{tabular}{cc}
\includegraphics[width=0.46\linewidth]{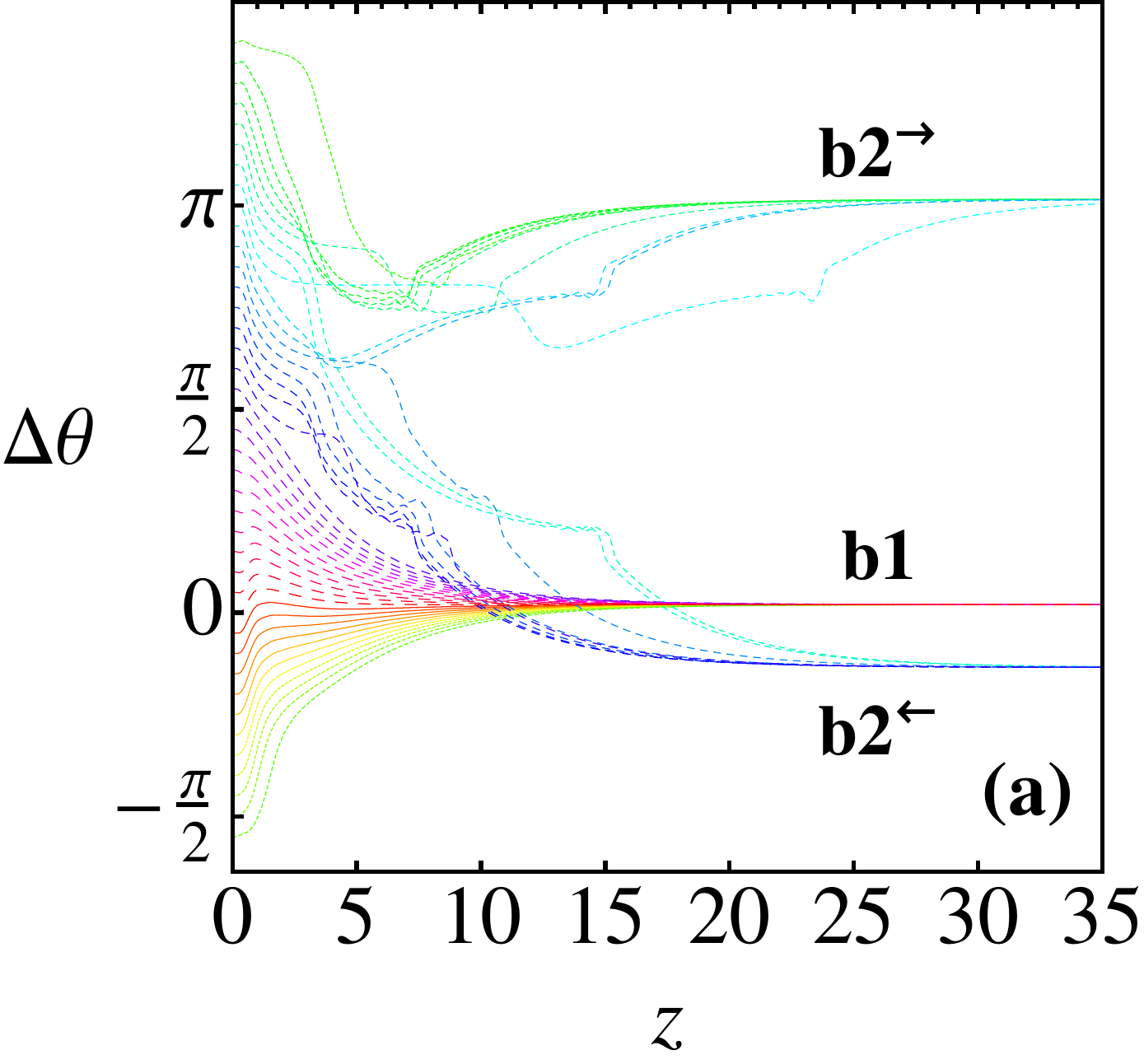}&
\includegraphics[width=0.46\linewidth]{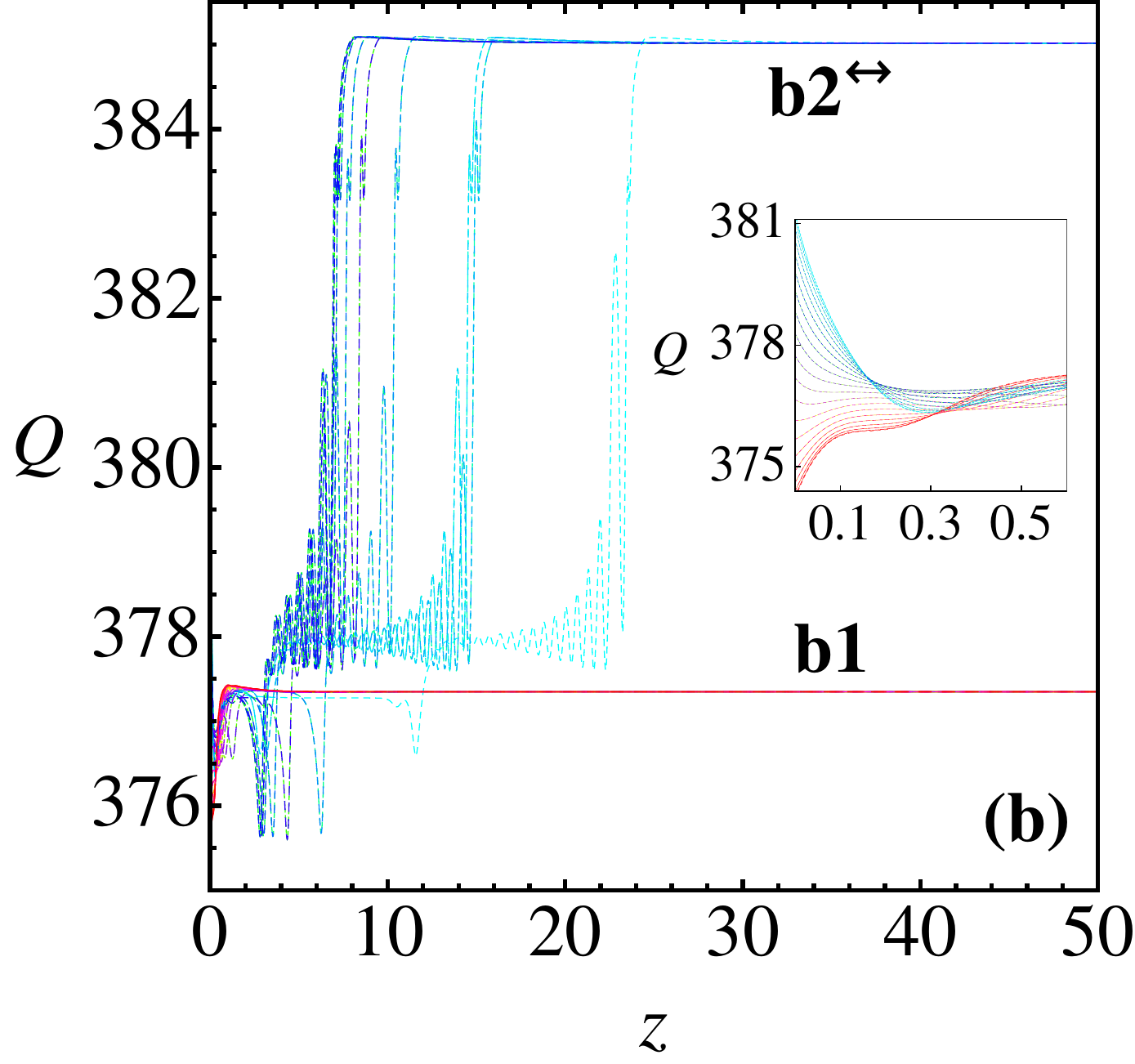}
  \end{tabular}
\caption{(Color online) (a) Dynamic evolution of the
    relative phase between two sites enclosed by the white circles, on
    the bound states showed in Fig.(\ref{fig7}). (b) optical power
    evolution for the same bound state. The inset in (b) shows a
    magnification of the initial stage of evolution.}
\label{fig6}
\end{figure}

\begin{figure}[ht] 
\centering 
\begin{tabular}{ccc}
\includegraphics[width=0.46\linewidth]{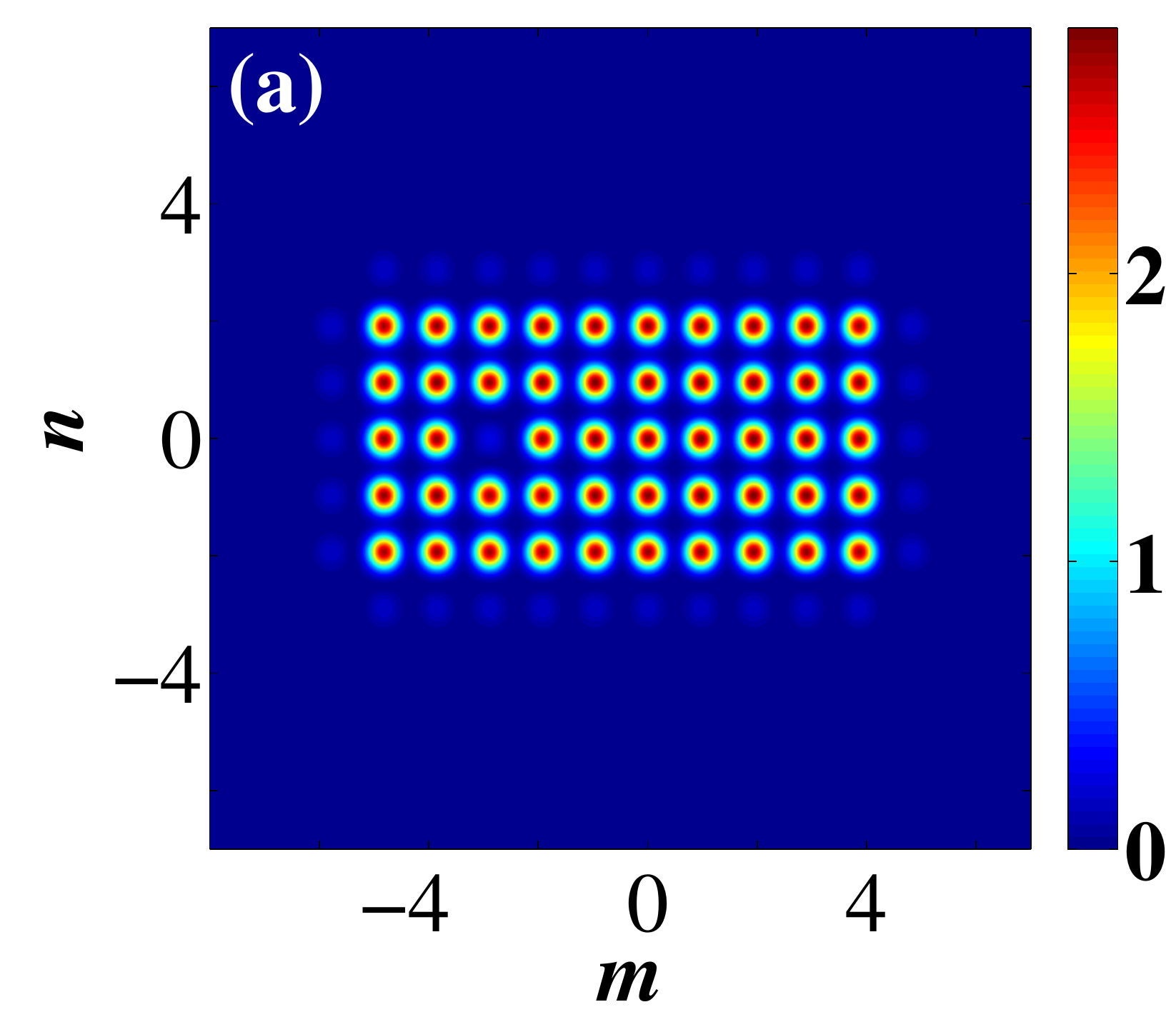}&
\includegraphics[width=0.475\linewidth]{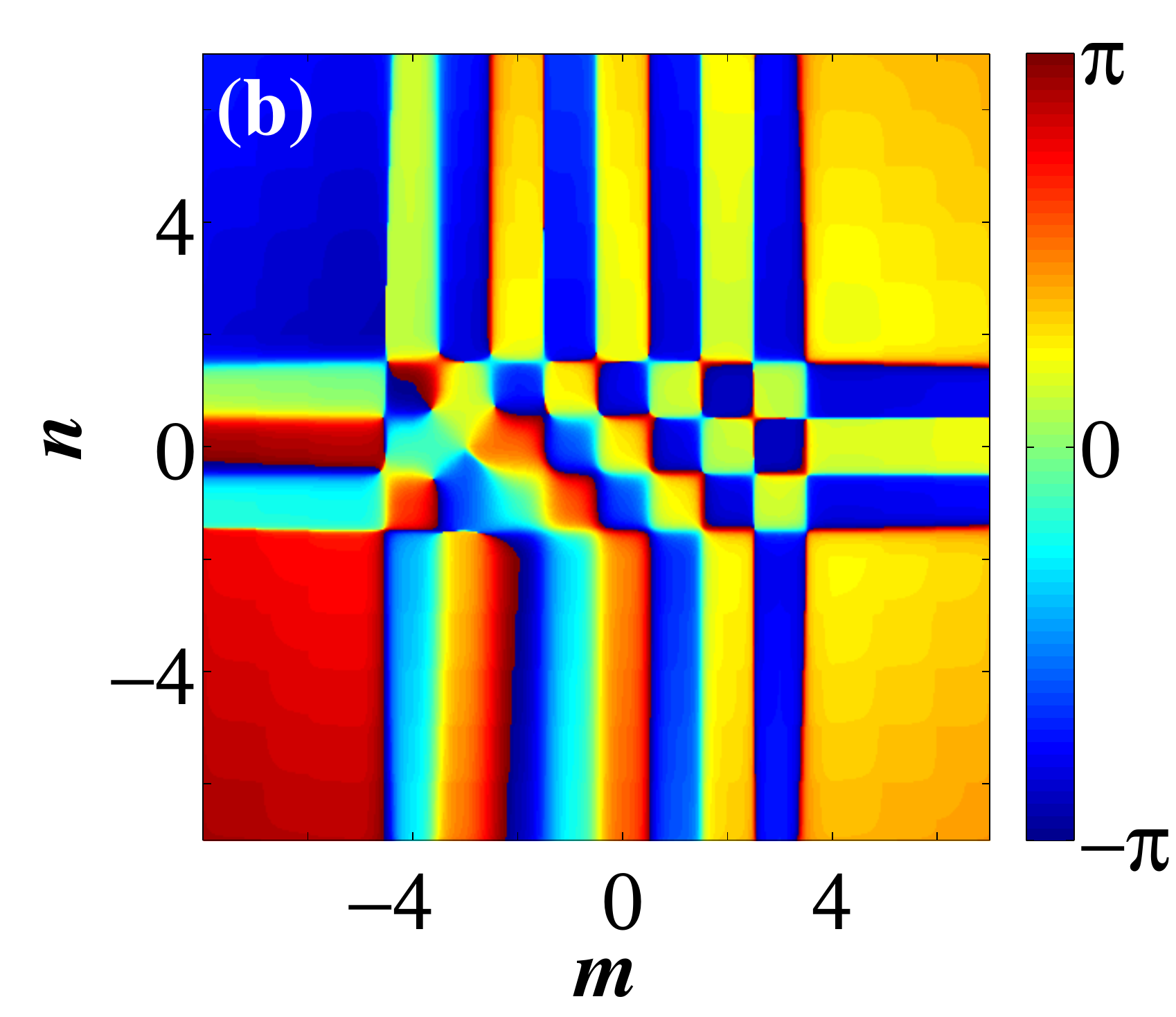}\\
\includegraphics[width=0.46\linewidth]{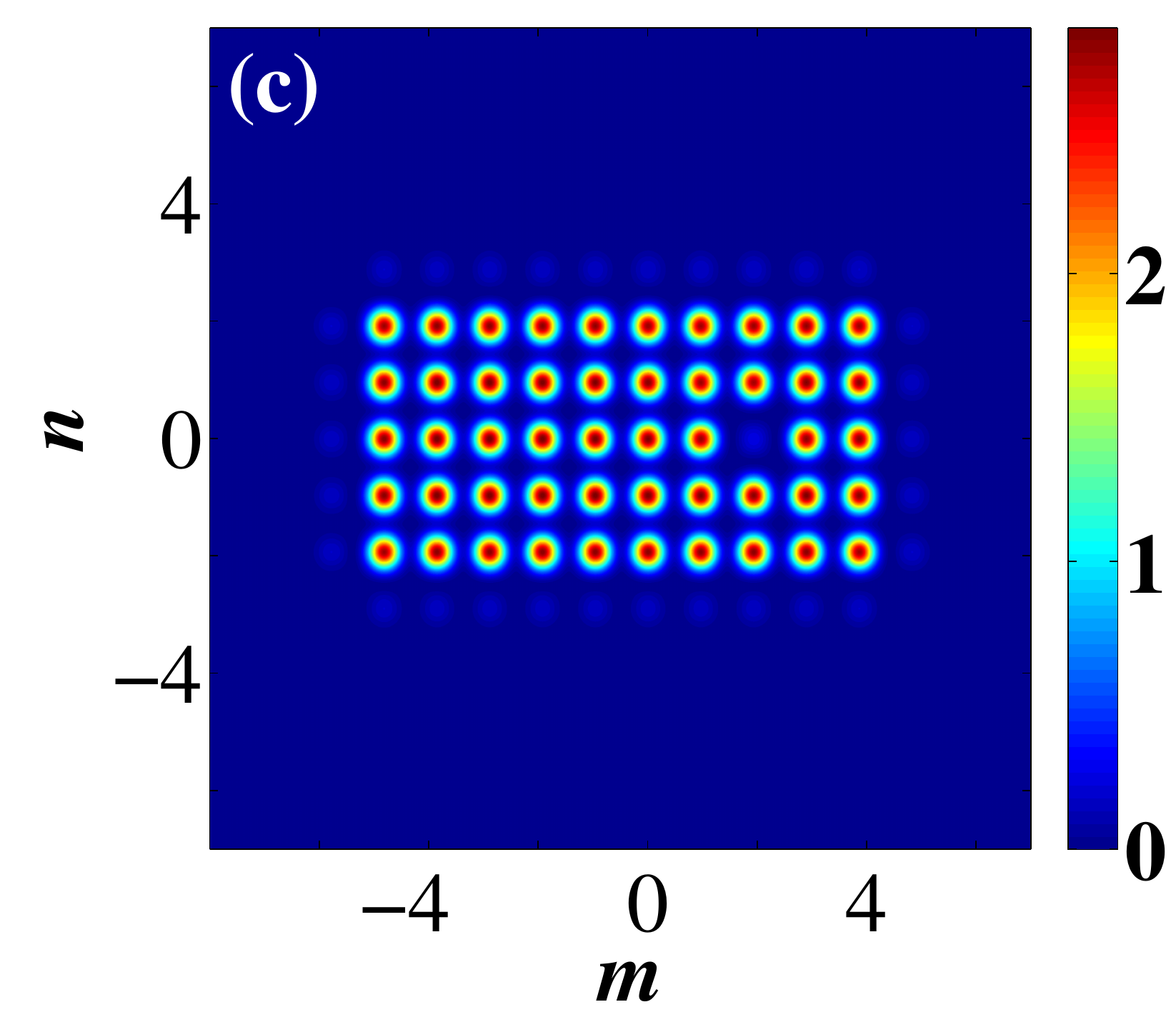}&
\includegraphics[width=0.475\linewidth]{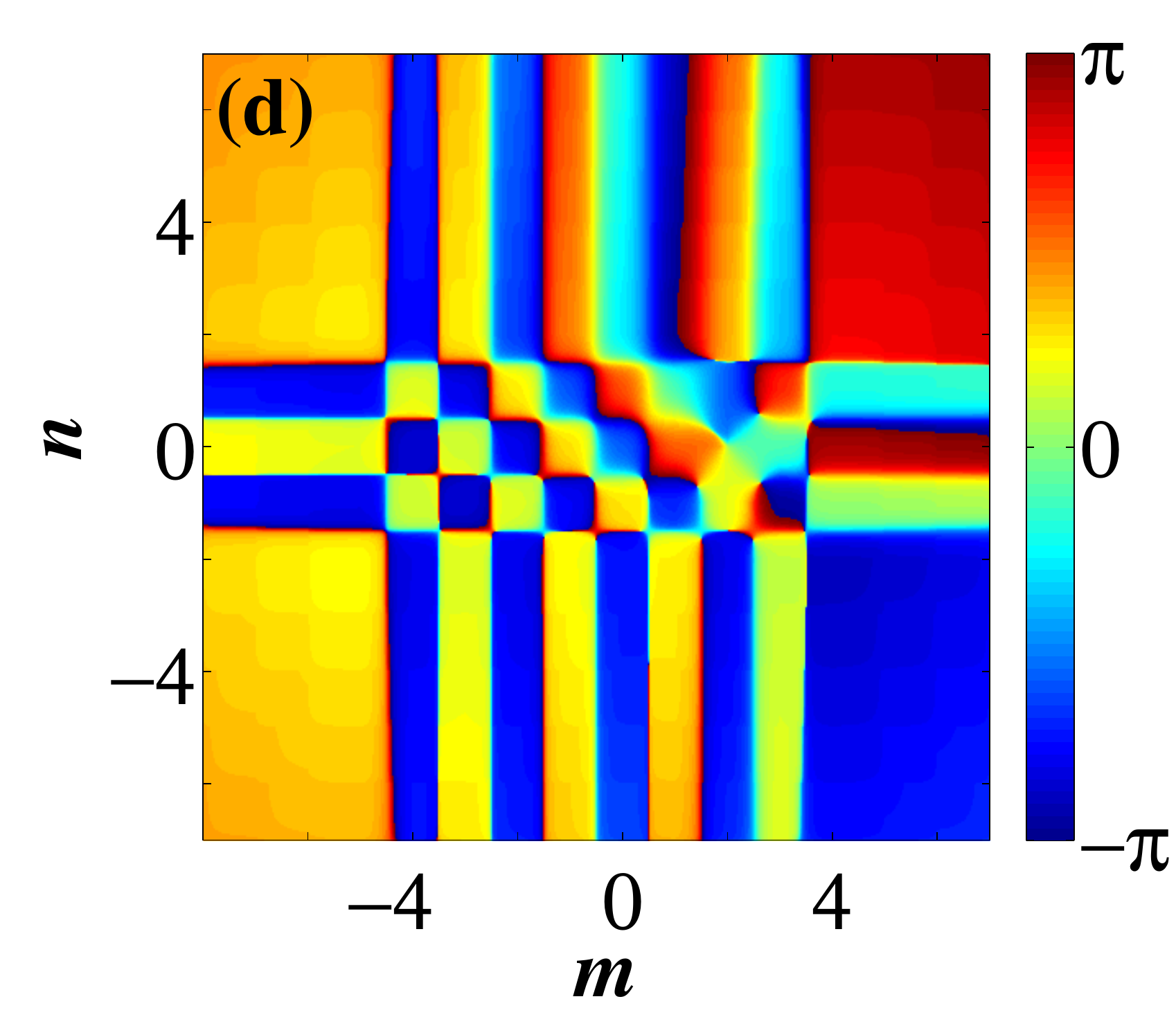}\\
\includegraphics[width=0.46\linewidth]{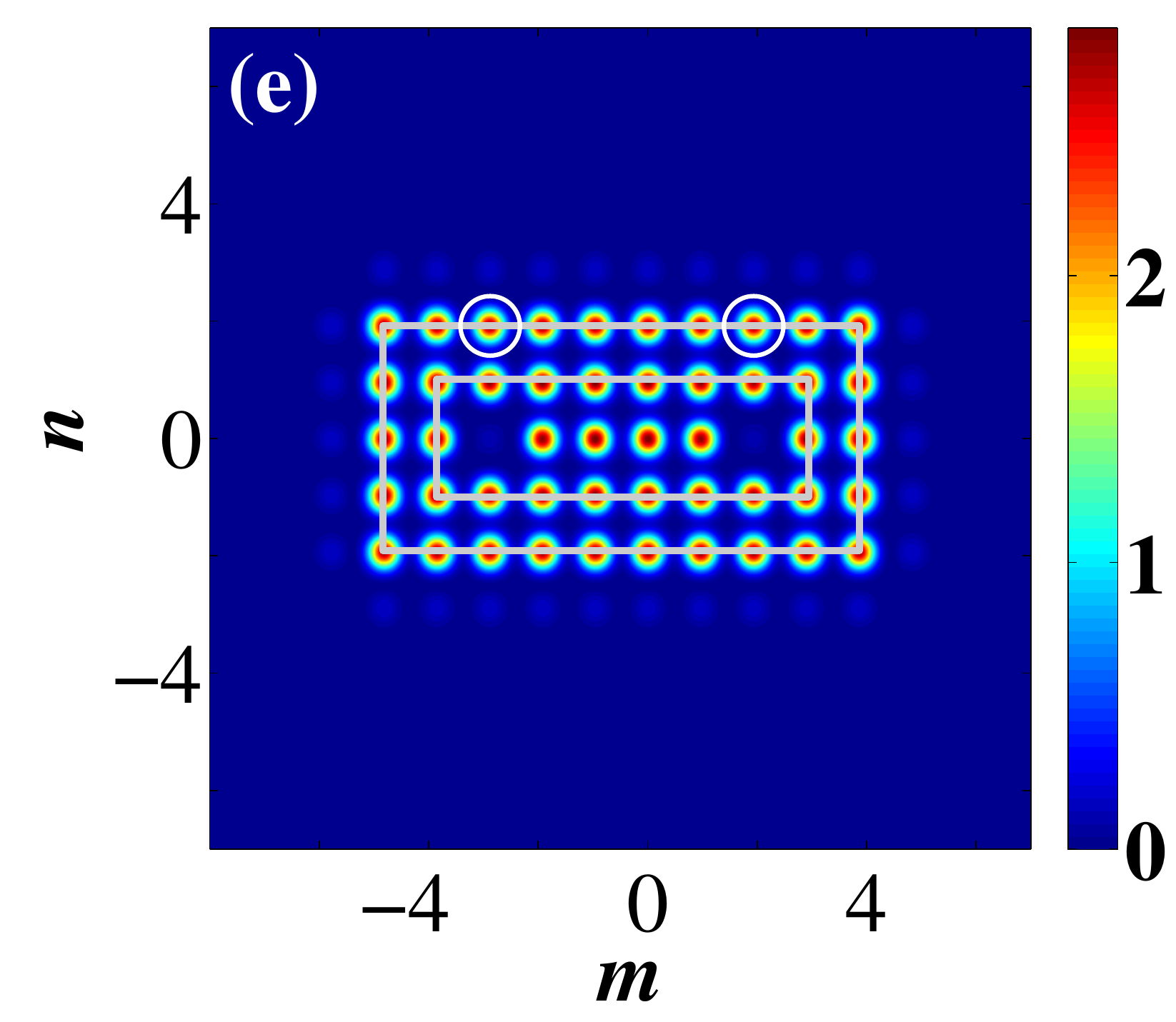}&
\includegraphics[width=0.475\linewidth]{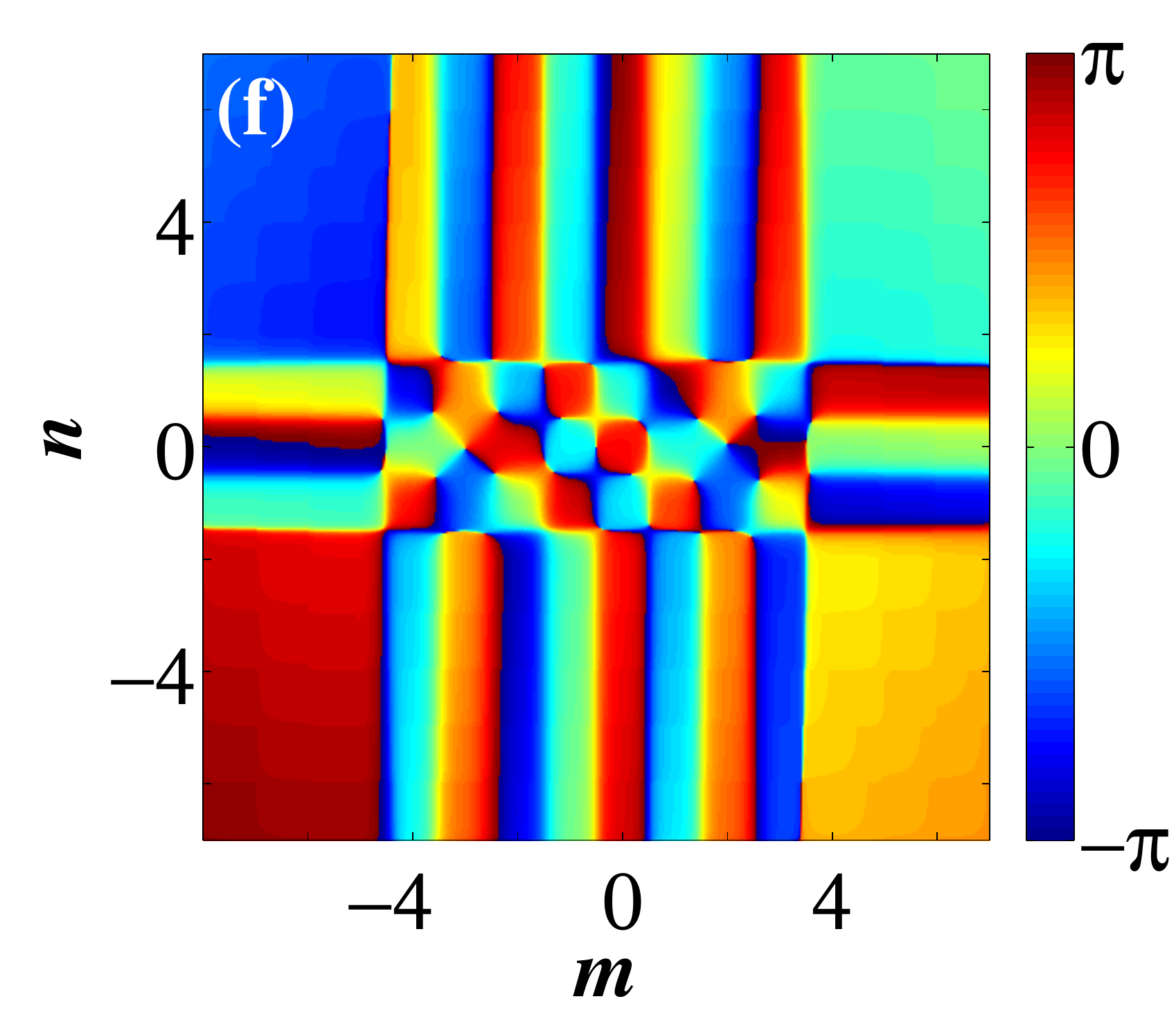}
  \end{tabular}
\caption{(Color online) Color map plots showing
    the amplitude (left) and phase (right) profiles for the stable
    solutions corresponding with the basins of attraction showed in
    Fig.\ref{fig6}. For basin {\bf b2$^{\leftarrow}$} the stable
    structure is similar to the profiles shown in (a) and (b), while (c)
    and (d) correspond with {\bf b2$^{\rightarrow}$}. For the remaining
    $b1$ basin, the vortex soliton is similar to the profiles shown in (e) and
    (f).} 
\label{fig7} 
\end{figure}

We consider now a second initial configuration where the two initially
independent {\bf E}-family vortices are placed closer to each other. 
We find a similar evolution than
before but now, there are three different stable attraction basins
for the relative phase evolution [see Fig.\ref{fig6}(a)]. Two of them,
the lowest ({\bf b2$^{\leftarrow}$}) and the highest ({\bf
  b1$^{\rightarrow}$}), correspond to the larger $Q$-value basin [{\bf
  b2$^{\leftrightarrow}$} in Fig.\ref{fig6}(b)]. The amplitude and
phase profiles for these two vortices solutions are shown in
Figs.\ref{fig7}(a) and (b), and (c) and (d), respectively. We can see
from the amplitude profiles that these solutions lost one of the two
original central cores (both solutions are equivalent if we perform a
inversion symmetry through the $\hat n$-axis). The global vorticity 
is lost, and we can see from Figs.\ref{fig7}(b) and (d) how the phase
circulation is truncated when we move to the region without a core.
Here, we claim that this mixed bound state is composed of an {\bf
  E}-family vortex soliton and a staggered bright soliton (with a
$\pi$-phase shift between nearest neighbors).

\begin{figure}[ht]
\center
\begin{tabular}{cc}
\includegraphics[width=0.6\linewidth]{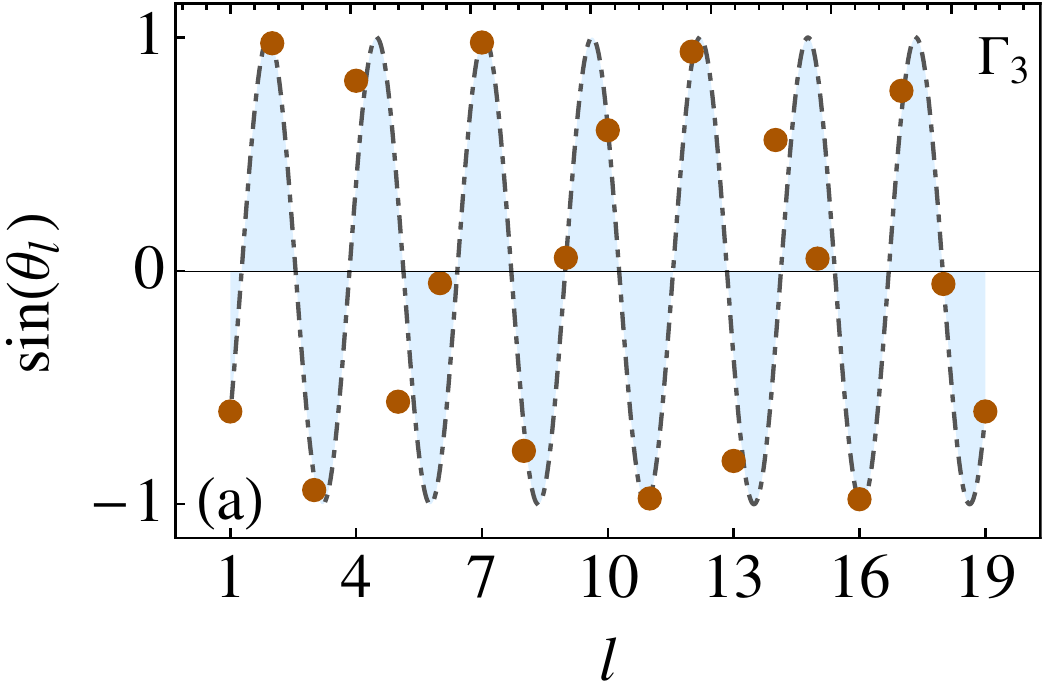}\\
\includegraphics[width=0.6\linewidth]{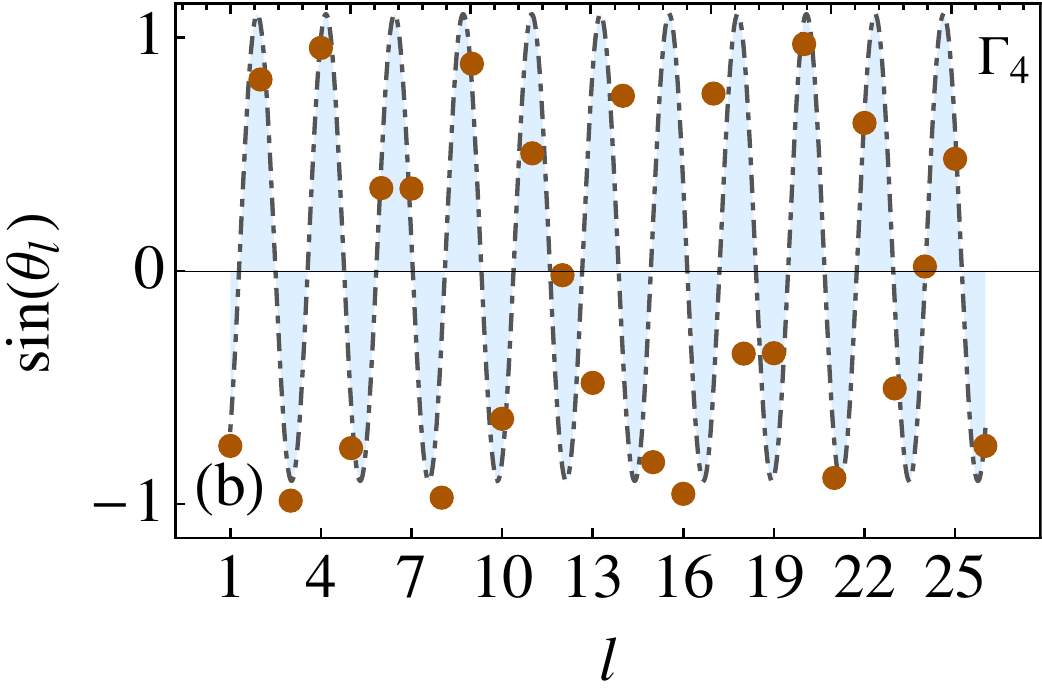}
\end{tabular}
\caption{(Color online) $\sin(\theta_{l})$ vs $l$ (site number) diagram for the
first (a) and second (b) discrete contours shown on the Fig.\ref{fig6}(a) for 
the bound state vortex soliton. }
\label{fig8}
\end{figure}

The third basin ({\bf b1}), which corresponds to the lower $Q$-value
basin in Fig.\ref{fig6}(b), has the amplitude and phase profiles
displayed in Figs.\ref{fig7}(e) and (f). We clearly see that it
preserves the initial two central cores keeping the same topological
charge as the initial condition. Unlike the previous two basins, the
global topological charge of this solution is well-defined. As for the
first configuration, there are also two different topological charges
for this composite vortex soliton. Again, to corroborate this, 
we plot $\sin(\theta_{l})$ vs $l$ in order to show in
detail the topological charge along two different contours. The first
one, $\Gamma_{3}$, corresponds to the inner rectangular contour
  sketched in Fig.\ref{fig7}(e), while $\Gamma_{4}$ corresponds with
  the outer rectangular contour sketched in the same figure.
Fig.\ref{fig8}(a) shows how the inner charge is $S=7$ while
Fig.\ref{fig8}(b) indicates a charge $S=11$ for contour $\Gamma_{4}$.

We show now two more examples of composite structures built from
an initial  superposition of two-and four solutions taken from the
{\bf D} and {\bf E} families shown in Fig.\ref{fig1}. In both cases,
the values of the parameters used are: $C=0.8$, $\delta=-0.9$,
$\varepsilon=0.9$, $\mu=-0.1$ and $\nu=0.1$. The typical propagation
distance was $z\approx 300$, enough for the power content to become constant.

\begin{figure}[ht]
\centering
\begin{tabular}{ccc}
\includegraphics[width=0.46\linewidth]{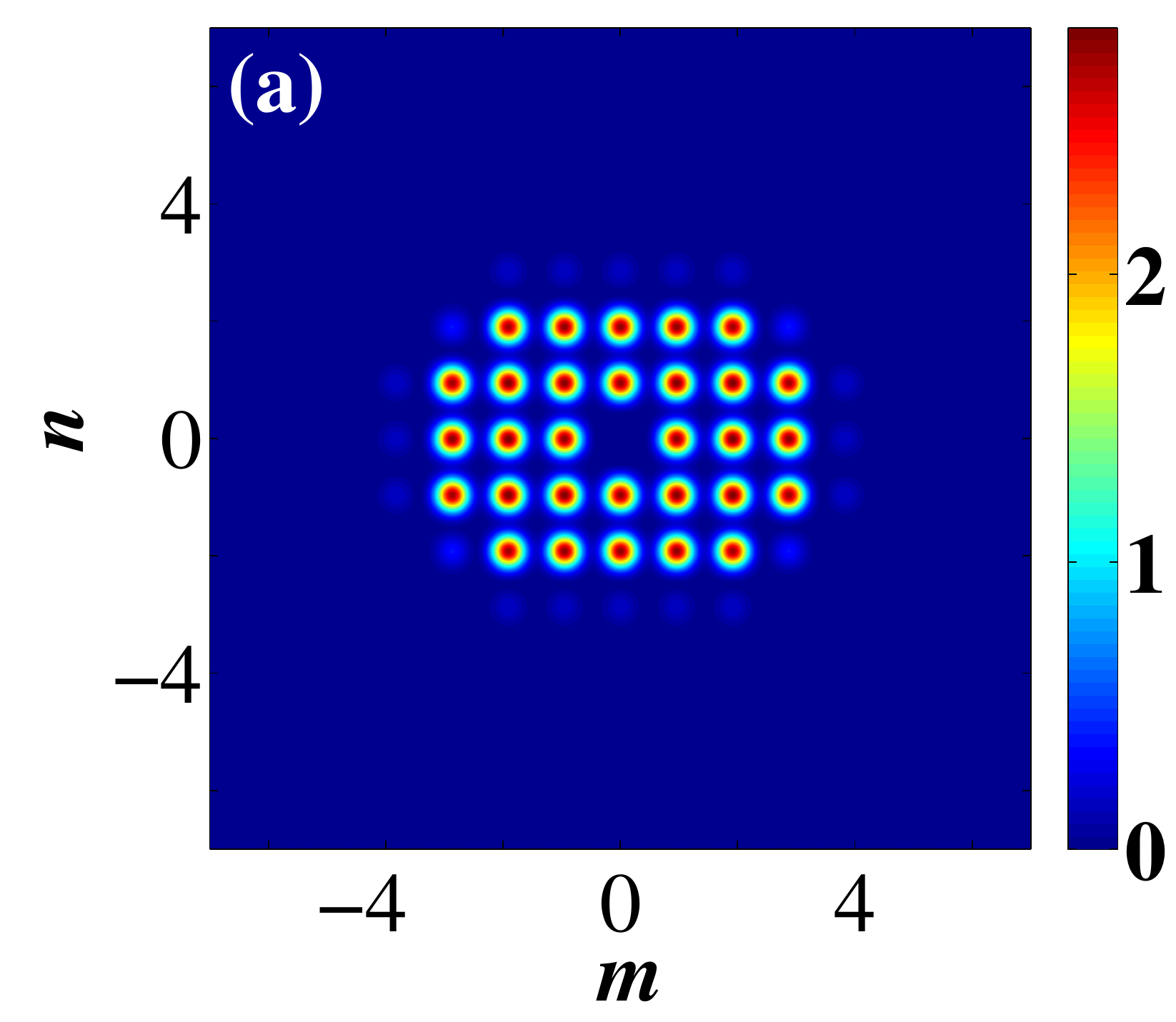}&
\includegraphics[width=0.475\linewidth]{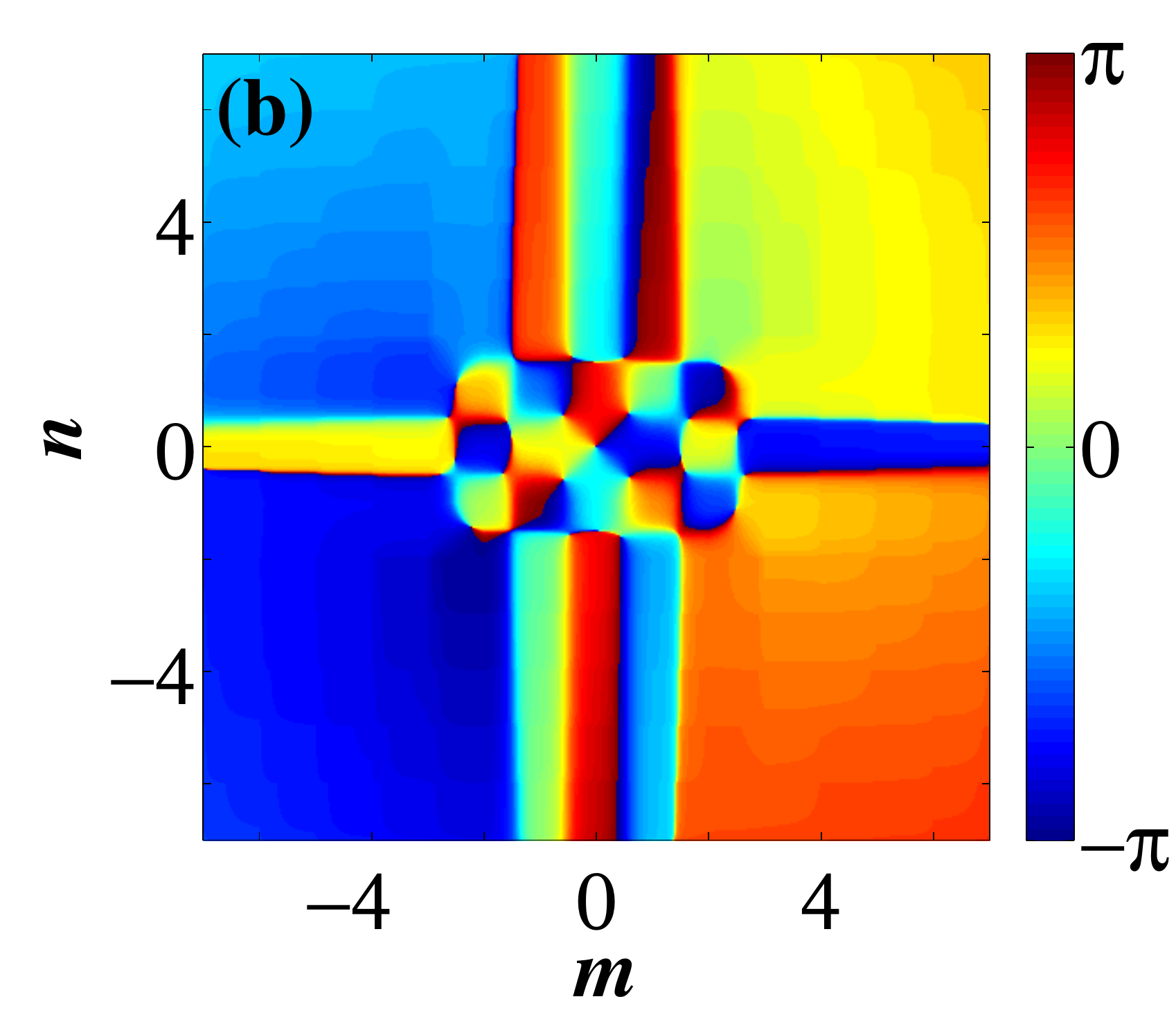}\\
\includegraphics[width=0.46\linewidth]{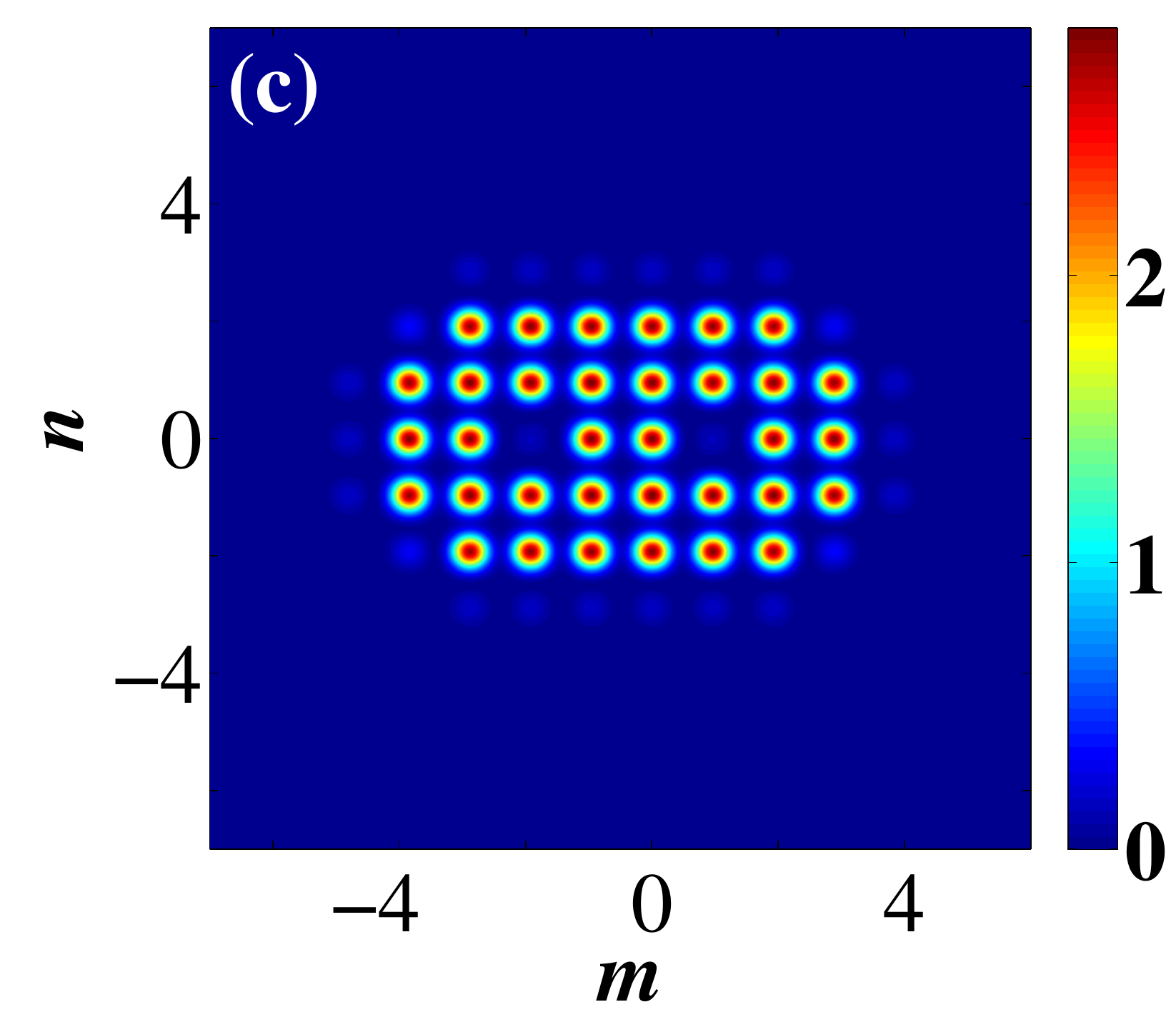}&
\includegraphics[width=0.475\linewidth]{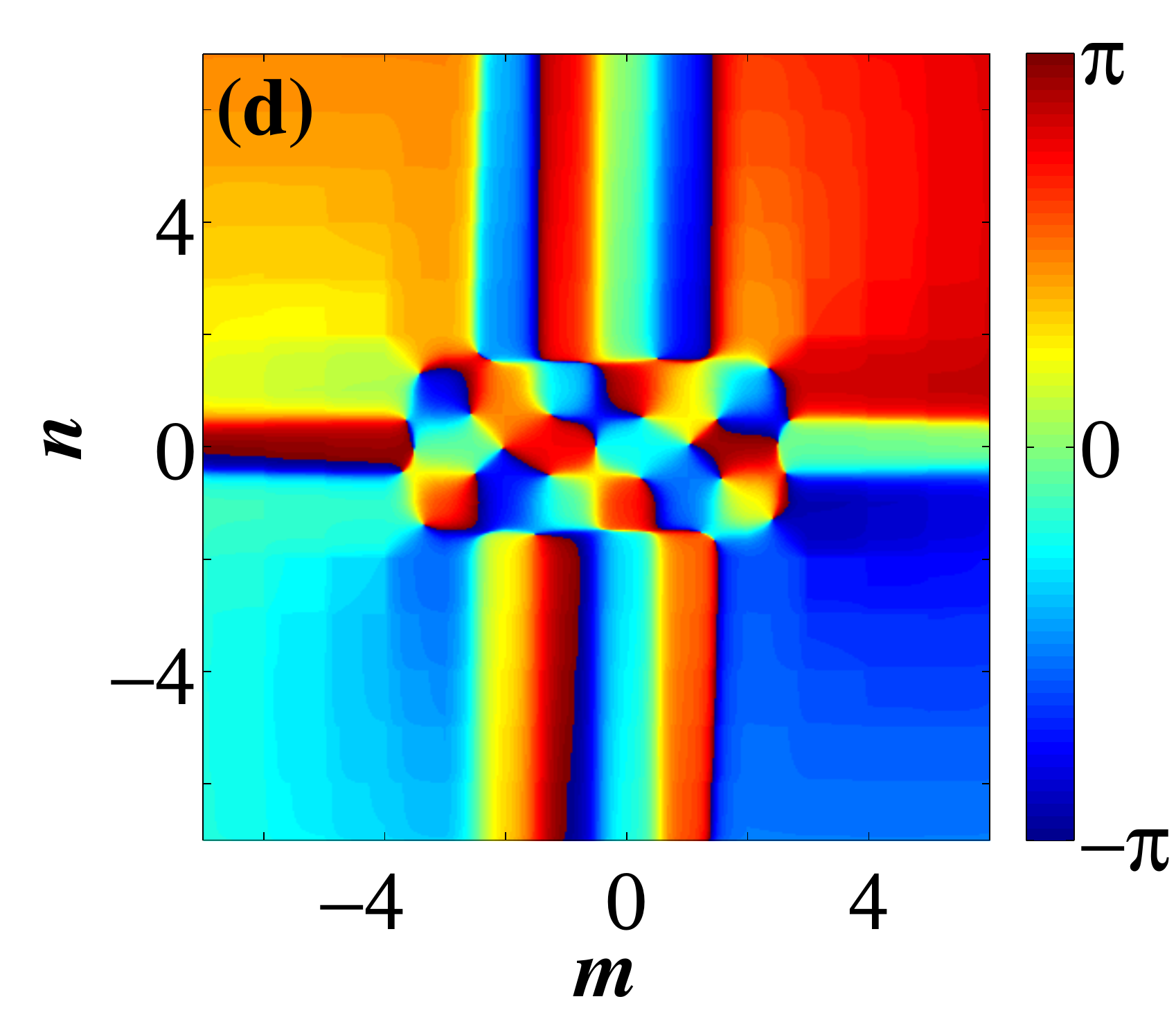}\\
\includegraphics[width=0.46\linewidth]{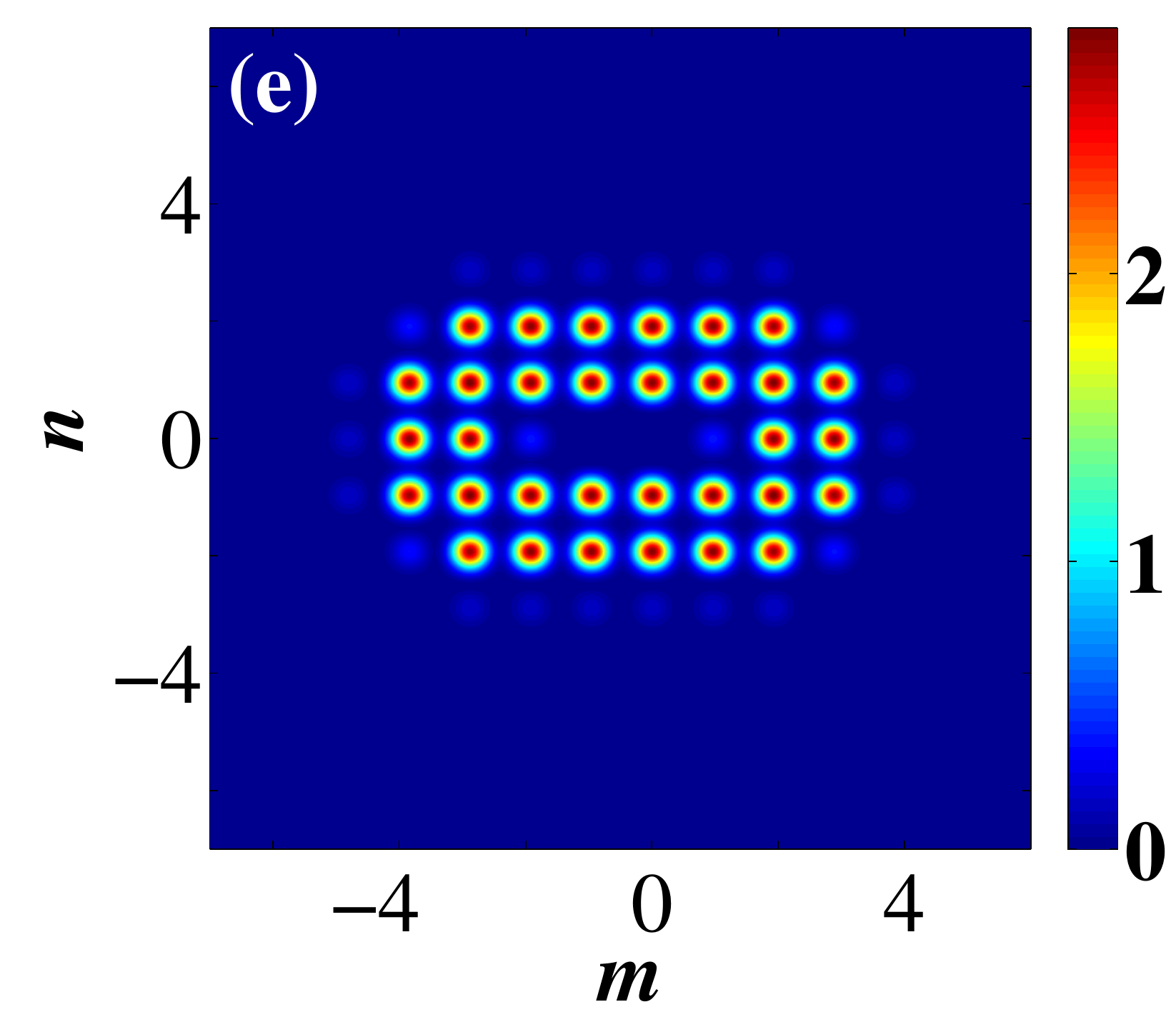}&
\includegraphics[width=0.475\linewidth]{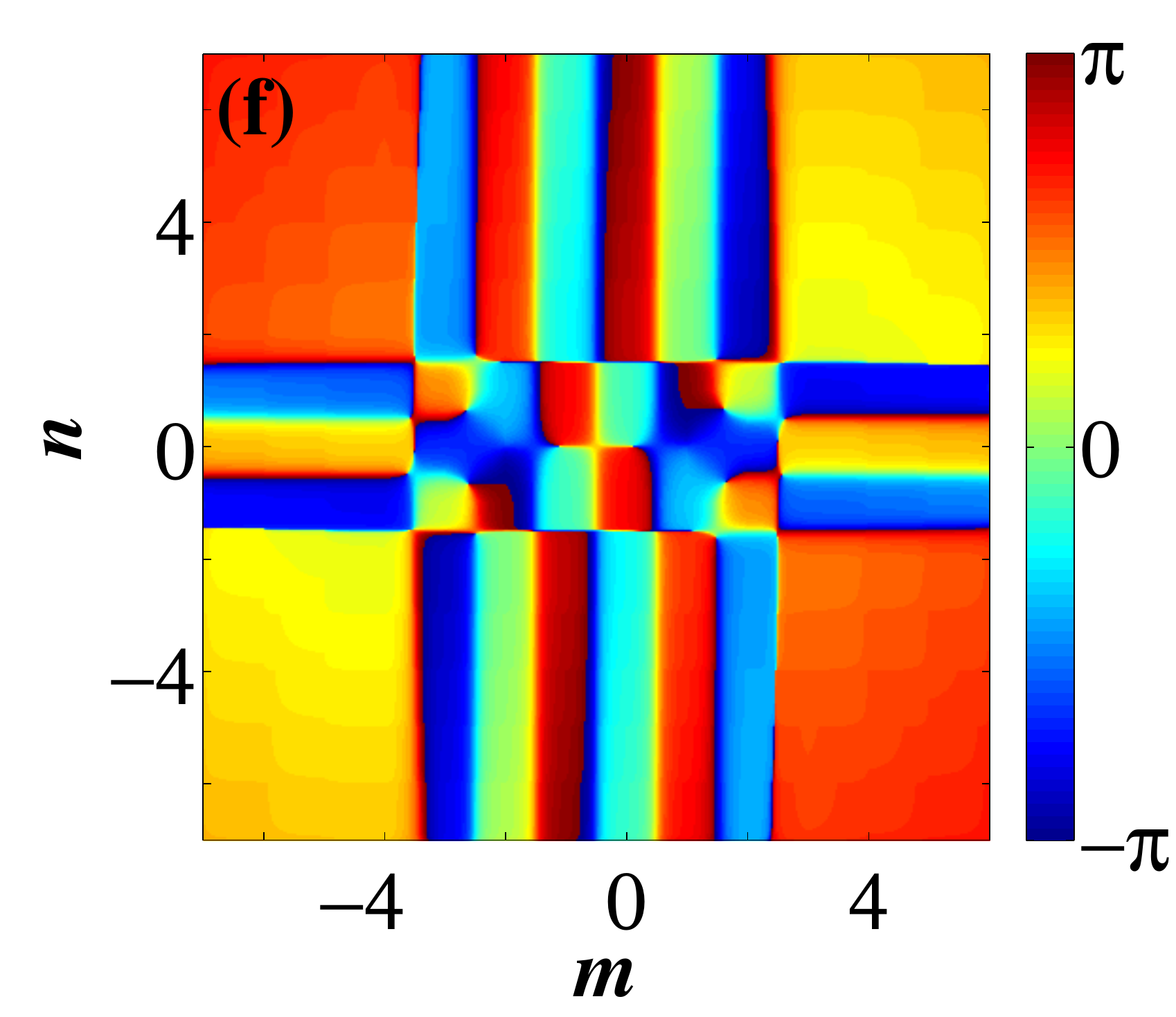}
\end{tabular}
\caption{(Color online) 
Color map plots showing the amplitude (left) and
phase (right) profiles for three stable vortex solutions 
obtained by superposing two vortex solitons belonging to the $D$ family,
at different initial distances.}
\label{fig9}
\end{figure}

Figure \ref{fig9} shows three stable solutions obtained by superposing
two vortex solitons belonging to the {\bf D}-family in Fig.\ref{fig1}.
The first one is constructed by overlapping two of these vortices
spaced by one site between their central cores. Figs.\ref{fig9}(a)
and (b) show the amplitude and phase profiles for this stable solution.
We note that this state has only one central core, located halfway
between the initial ones. On the other hand, the phase profile shows
a charge $S=5$ at the inner contour and rotated with respect to the next
discrete contours.

The next configuration is constructed in the same manner as the
previous one but now, the center-to-center distance between the cores has
been increased to two sites. Figure~\ref{fig9}(c) shows a dynamically
stable solution with two central cores located at the same positions as
the initial condition. The phase profile [see Fig.\ref{fig9}(d)] shows
a value of $S=5$ for the topological charge, as in the previous case. 
A third stable composed structure is obtained by superposing again two
vortex solitons with an initial center-to-center distance of three sites.
The amplitude profile for the new dynamically stable solution has one
horizontally elongated core, along two lattice sites, as shown in Fig.\ref{fig9}(e),
and its topological charge has two different values as shown in
Fig.\ref{fig9}(f). Indeed, the innermost  discrete contour exhibits a
charge $S=6$, while the remaining contours have a charge $S=10$.

Finally, we show another example of a composed structure that was
obtained by combining four solutions belonging to the {\bf E} family .
We locate each {\bf E}-vortex by placing their central cores forming the
vertices of a $8\times 8$ square. We use this configuration as
initial condition for model (\ref{dgl2}) and find a dynamically stable
stationary solution.
\begin{figure}[ht]
\centering
\begin{tabular}{cc}
\includegraphics[width=0.46\linewidth]{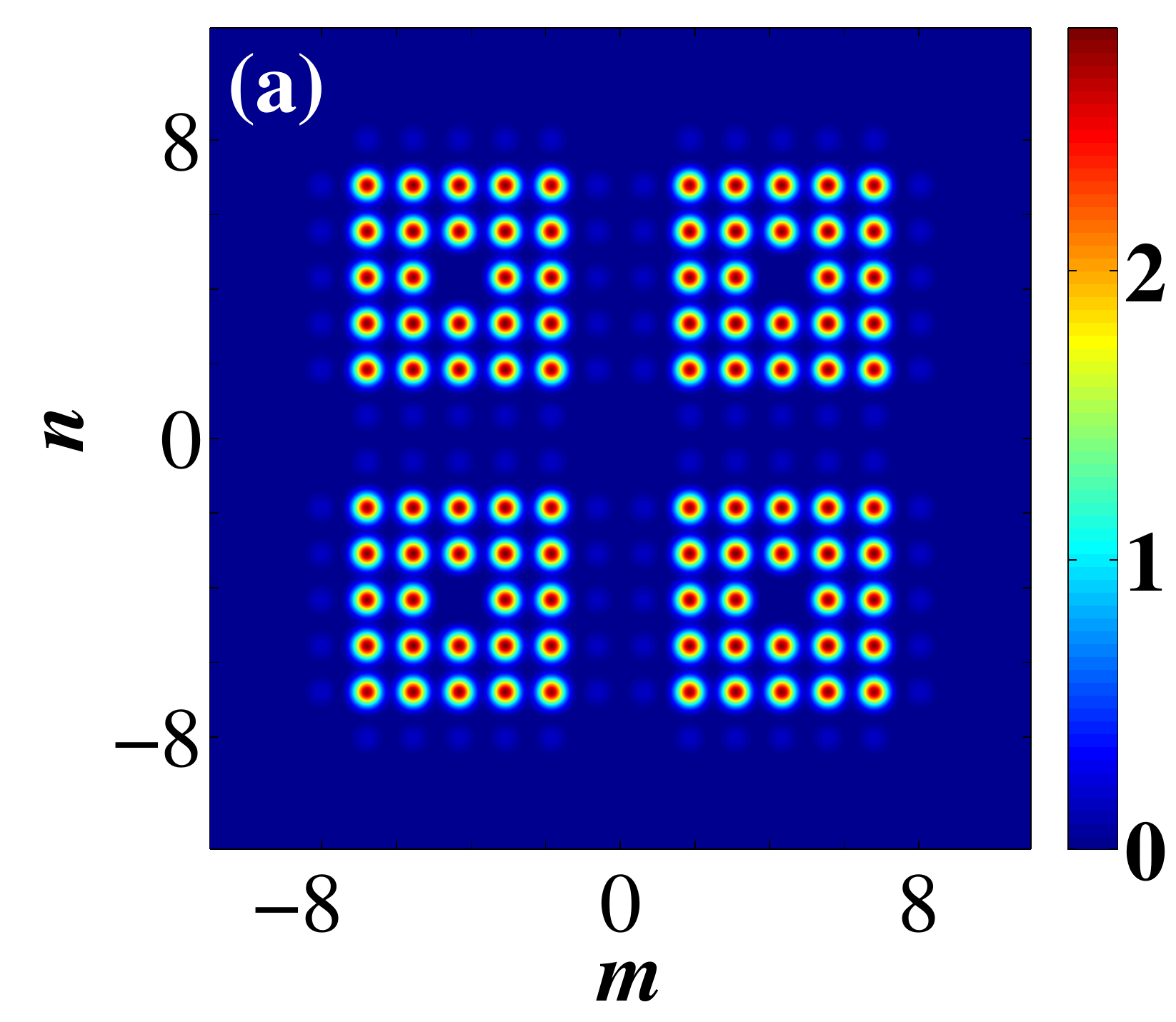}&
\includegraphics[width=0.47\linewidth]{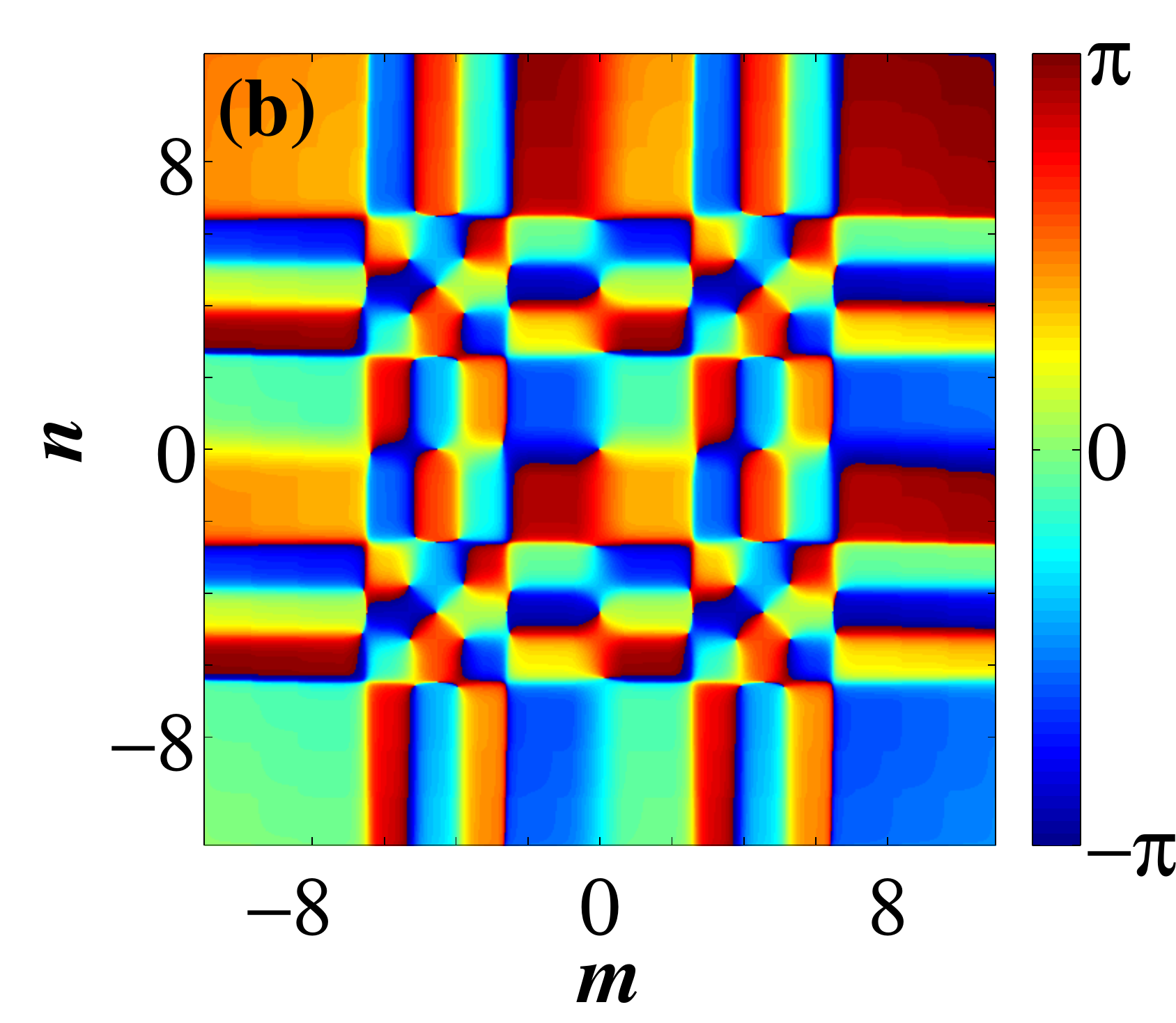}
\end{tabular}
\caption{(Color online) Color map plots for the amplitude (a) and phase
(b) profiles for a dynamically stable
solution obtained by combining four solutions belonging to the {\bf E} family}. 
\label{fig10}
\end{figure}
Figs.~\ref{fig10}(a) and (b) show the amplitude and phase profiles for
this composite solution. We observe a spatial amplitude distribution
similar to the initial condition, where the four initial cores preserve
their initial position and vorticity. In addition, an extra phase core
appears at the lattice center ($n,m=0$), around which a $S=-1$ topological
charge can be observed. If we follow a new contour that encloses all
the sites with large amplitude, the topological charge measured will
be $S=15$.

\section{Schr\"odinger limit}
\label{schr}

Most of the present experiments with optical beams are performed under
conditions that closely meet the cubic conservative limit. So, we are
interested in knowing if all these previous dissipative structures
can be observed also here. In this scenario, all the dissipative
parameters are suppressed, i.e., $\nu=\mu=\varepsilon=\delta=0$,
and model (\ref{adgl2}) reduces to the discrete NLSE equation
\begin{equation}
-\lambda \psi_{m,n}+\hat C\psi_{m,n}+|\psi_{m,n}|^{2}\psi_{m,n}=0.
\label{dnls}
\end{equation}
Taking as an \emph{ansatz} for the decoupled limit for some of the
solutions previously described, we have obtained the same kind of
structures for the discrete NLSE. We have constructed the corresponding
families and also computed their corresponding stability region.
We note that all these solutions exist in the Schr\"odinger limit,
but are stable only for high values of their optical power content.

\begin{figure}
\centering
\includegraphics[width=0.95\linewidth]{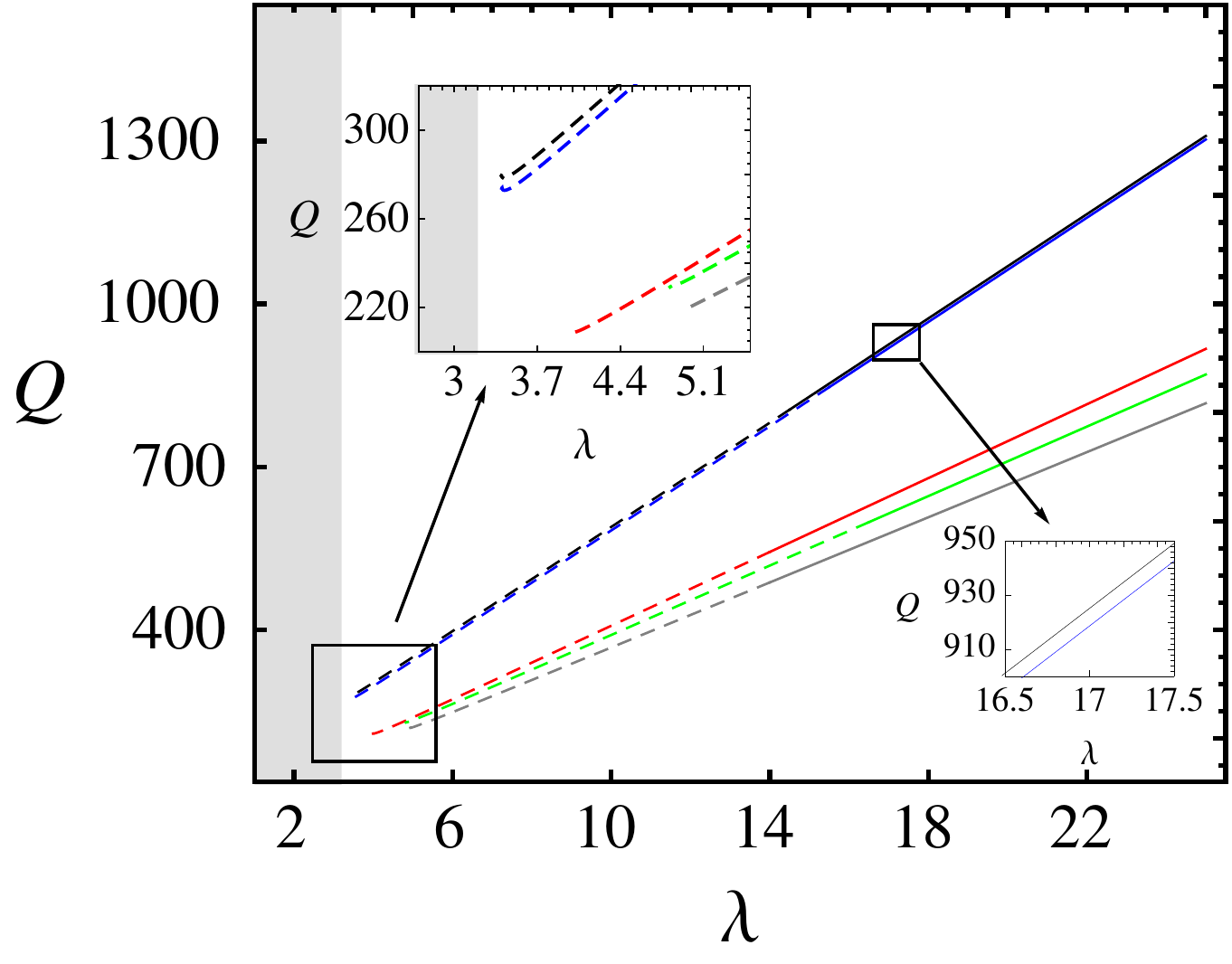}
\caption{$Q$ versus $\lambda$ diagram displaying the families
corresponding with some of dissipative solutions reported
previously, but for the conservative cubic case.}
\label{fig11}
\end{figure}

In Fig.(\ref{fig11}) we show five families of stationary solutions
of the NLSE that correspond with some of the solutions reported in the previous
sections. Namely, the blue line is the family corresponding with the
solution displayed in Fig.(\ref{fig3}a-b), the black line with the
Fig.(\ref{fig7}e-d), and the red, green and gray line with
Fig.(\ref{fig9}a-b), Fig.(\ref{fig9}c-d) and Fig.(\ref{fig9}e-f)
respectively. The stable (unstable) solutions are represented by
continuous (dashed) lines.
The inset located in the upper left corner at Fig.(\ref{fig11}) shows
a magnified view of the region close to the linear band (gray zone). As 
usual~\cite{PhysRevA.82.063818}, each one of these families tends to increase its power steeply after
passing through the point of minimum optical power. The other inset at the
lower right corner shows a zoom of the black and blue curves, which are
almost indistinguishable because they have very similar spatial profiles.

\section{Summary and conclusions}
\label{con}
We have reported several new families of discrete vortex solitons, characterized
for having two topological charges simultaneously, and coexisting 
for the same set of parameters. By superposing two or more of these
vortices, we have been able to produce new, dynamically stable
composite vortex solitons that are also endowed with multiple vorticity
charges. Additionally, we have shown that these composite structures persist 
in the conservative limit and they are stable for high values of their 
power content. 
 
\section{Acknowledgments}
C.M.C. and J.M.S.C. acknowledge support from the Ministerio de Ciencia
e Innovaci\'on under contracts FIS2006-03376 and FIS2009-09895. R.A.V
and M.I.M. acknowledge support from Fondecyt Grants 1110142 and
1080374, and Programa de Financiamiento Basal de Conicyt
(FB0824/2008).

\end{document}